\documentclass[12pt]{article}
\usepackage{putex}
\usepackage{amsmath,amssymb}
\usepackage{graphicx}
\usepackage{xcolor}
\usepackage{bm}
\usepackage{cite}
\usepackage{slashed}
\usepackage{tikz-cd}
\usepackage[backref]{hyperref}

\graphicspath{{./Figures/}}

\newcommand{\bZ}{\ensuremath{\mathbb{Z}}}

\newcommand{\es}[2] {\begin{equation} \label{#1} \begin{split} #2 \end{split} \end{equation}}

\newcommand{\bsigma}{\bm{\sigma}}

\newcommand{\spc}[2][c]{\begin{tabular}[#1]{@{}c@{}}#2\end{tabular}}

\def\U{\mathrm{U}}
\def\SU{\mathrm{SU}}
\def\SO{\mathrm{SO}}
\def\USp{\mathrm{USp}}
\renewcommand{\O}{\mathrm{O}}

\begin{document}

\pagenumbering{Alph}
\title{Gauging and Decoupling in 3d $\mathcal{N}=2$ dualities}

\authors{Jeongseog Lee$^{1}$ and Masahito Yamazaki$^{2,3}$}

\institution{PU}{\centerline{$^1$Department of Physics, Princeton University, Princeton, NJ 08544, USA}}
\institution{IPMU}{\centerline{$^2$Kavli IPMU (WPI), University of Tokyo, Kashiwa, Chiba 277-8583, Japan}}
\institution{IAS}{\centerline{$^3$Institute for Advanced Study, School of Natural Sciences, Princeton, NJ 08540, USA}}

\abstract{
One interesting feature of 3d $\mathcal{N}=2$ theories 
is that gauge-invariant operators can decouple 
by strong-coupling effects, leading to emergent flavor symmetries in the IR.
The details of such decoupling, however, depends very delicately on the 
gauge group and matter content of the theory.
We here systematically study the IR behavior of 3d $\mathcal{N}=2$ SQCD with 
$N_f$ flavors, for gauge groups $\SU(N_c), \USp(2N_c)$ and $\SO(N_c)$.
We apply a combination of analytical and numerical methods, both to
small values of
 $N_c, N_f$ and also to the Veneziano limit, where $N_c$ and $N_f$ are taken to be large with their ratio $N_f/N_c$ fixed.
We highlight the role of the monopole operators and the interplay with Aharony-type dualities.
We also discuss the effect of gauging continuous as well as discrete flavor symmetries,
and the implications of our analysis to the classification of $1/4$--BPS co-dimension 2 defects of 6d $(2,0)$ theories.
}

\preprint{IPMU-16-0011}

\maketitle

\pagenumbering{arabic}
\setcounter{page}{1}

\tableofcontents

\section{Introduction and Summary}

In this paper we study three-dimensional $\mathcal{N}=2$ supersymmetric gauge theories \cite{Aharony:1997bx,deBoer:1997kr}. Since the gauge coupling is dimensionful in three spacetime dimensions, we expect
that generic three-dimensional gauge theories become strongly-coupled in the deep IR (infrared),
where the non-perturbative effects play prominent roles.
For example, in three-dimensional $\mathcal{N}=2$ pure $\SU(N_c)$ super Yang-Mills theory 
non-perturbative instanton effects generate a superpotential term, which lifts the supersymmetric vacuum \cite{Affleck:1982as}.
 
One interesting feature of the IR behavior of 3d $\mathcal{N}=2$ supersymmetric gauge theories is that 
there are often indications that 
strong-coupling effects make some operators free, and decouple those from the rest of the system,
in the IR. In this case we need to subtract the corresponding degrees of freedom to 
discuss truly strongly-coupled interacting dynamics.
This also means that there are emergent $\U(1)$ flavor symmetries in the IR, which act only on that decoupled fields.

That some operators could decouple in the IR is known also in four dimensions, e.g.\ from the analysis of the 
4d $\mathcal{N}=1$ adjoint QCD \cite{Kutasov:1995ve,Kutasov:1995np}. 
The story is, however, even richer in the three-dimensional counterparts discussed in this paper. This is because in three dimensions we have monopole operators (constructed out of dual photons), 
which are new sources for possible IR decouplings.
Indeed, we will see below a strong evidence that such a decoupling of monopole operators do happen for 
3d $\mathcal{N}=2$ $\SU(N_c)$ gauge theory with $N_f$ flavors, for infinitely many values of $N_c$ and $N_f$
(see \cite{Safdi:2012re} for a similar analysis for $\U(N_c)$ gauge groups, which provided an inspiration for this paper).
This is in contrast with their 4d $\mathcal{N}=1$ counterparts, which show no sign of such decouplings.

One useful signal of the IR decoupling of operators is the unitarity bound \cite{Mack:1975je,Flato:1983te,Dobrev:1985qv,Minwalla:1997ka} (see \cite{Penedones:2015aga,Yamazaki:2016vqi} for recent discussion).
In 3d $\mathcal{N}=4$ theories there is a simple formula for the
scaling dimensions of the monopole operators \cite{Borokhov:2002cg}, 
which lead to the good/ugly/bad classifications of 3d $\mathcal{N}=4$ theories \cite{Gaiotto:2008ak}.
More concretely, the absence of the IR decouplings for $\U(N_c)$ 3d $\mathcal{N}=4$ SQCD (Supersymmetric QCD) with $N_f$ flavors require a simple inequality $N_f> 2N_c$.

One natural question is then what happens to the case of reduced supersymmetry, i.e.\ 3d $\mathcal{N}=2$ supersymmetry.
In this paper we study 3d $\mathcal{N}=2$ SQCD with 
$N_f$ flavors.\footnote{In this paper we only discuss parity-preserving theories, and in particular we do not discuss theories with Chern-Simons terms.
The Chern-Simons terms renders the monopole operator to be gauge variant, which significantly modifies the discussion below, 
as already commented in \cite{Safdi:2012re}.}
In this case, the conformal dimension of a chiral primary operator (such as the monopole operator) is determined by its R-charge. The complication is that the UV (ultraviolet) $\U(1)$ R-symmetry
could mix in the IR with flavor $\U(1)$ symmetries, hence the IR $\U(1)$ R-symmetry in the IR superconformal algebra is in general different from the UV R-symmetry.

The correct IR R-symmetry can be determined with the help of the $F$-maximization \cite{Jafferis:2010un},
i.e.\ the maximization of the supersymmetric partition function on the round three-sphere $S^3$ \cite{Kapustin:2009kz,Jafferis:2010un,Hama:2010av}. However, the $S^3$ partition function is a complicated integral expression,
an evaluation of which often requires numerical analysis, and it turns out that whether or not the IR decoupling happens or not 
depends very sensitively on the choice of the gauge groups and the matter contents of the theory.
For this reason we consider SQCD with various different gauge groups,
$\SU(N_c), \USp(2N_c)$ and $\SO(N_c)$, with $N_f$ flavors,
with different values of $N_c$ and $N_f$.
Our analysis simplifies somewhat in the Veneziano limit
\es{Veneziano}{
N_f,  N_c \to \infty \ ,  \quad x:=\frac{N_f}{N_c} \quad \textrm{kept finite} \ .
}
In all the cases we find that there is a critical value $x_c>1$, below which some of the monopole operators decouple. 
Once some operators decouple we can re-do the $F$-maximization following the prescription of \cite{Niarchos:2011sn,Morita:2011cs,Agarwal:2012wd}.

When we carry out the $F$-maximization, we run into another subtlety:
the $S^3$ partition function does not always converge.
This causes a problem, since we need $S^3$ partition function to 
determine the correct IR scaling dimension.

What saves the day is that 3d $\mathcal{N}=2$ SQCD
has a non-perturbative magnetic dual, found by
Aharony \cite{Aharony:1997gp}  (see also \cite{Aharony:2013dha,Aharony:2013kma,Park:2013wta}).\footnote{The case of $\USp(2)$ 
gauge group is special, since $\SU(2)$ is also $\USp(2)$,
which is also the same as $\SO(3)$ up to global $\mathbb{Z}_2$ quotient.
This means that a single electric theory has several different magnetic duals.
}
Whenever the electric theory has a divergent partition function,
the magnetic partition function is convergent, 
whose $S^3$ partition function can
be used for $F$-maximization. 

We point out that a gauging of flavor symmetries (either continuous or discrete)
can drastically modify the IR behavior of the theory. We discuss this phenomena
for the following three examples (see later sections for precise notations):
\begin{itemize}
\item A gauging of the $U(1)_B$ symmetry of $\SU(N_c)$ SQCD, to obtain $\U(N_c)$ SQCD.
For the $\U(N_c)$ gauge group, the critical value $x_c$ in the Veneziano limit is
also the value where we switch from the electric to magnetic descriptions (this is also the case for $\USp(2N_c)$ and $\SO(N_c)$ gauge groups).
This is in contrast with the case of $\SU(N_c)$ SQCD,
where the magnetic description turns out to be valid above $x_c$ as well as below. 

\item A gauging of $\mathbb{Z}_2$ or $\mathbb{Z}_2\times \mathbb{Z}_2$ discrete flavor symmetries of the $\SO(N_c)$ SQCD, to 
obtain $\O(N_c)_{+}$, $\O(N_c)_{-}$, ${\rm Spin}(N_c)$ or ${\rm Pin}(N_c)$ SQCD.
Such a gauging changes the monopole operator with the minimal charge.
We find some examples where the monopole operator decouples for 
$\SO(N_c)$ and $\O(N_c)_{+}$ gauge groups, but not for 
$\O(N_c)_{-}$, ${\rm Spin}(N_c)$, ${\rm Pin}(N_c)$ gauge groups.

\item A gauging of $SU(N_f)_V$ flavor symmetry of $\U(N_c)$ SQCD, to 
obtain a quiver gauge theory with gauge group $\U(N_c)
\times \SU(N_f) \simeq (\U(N_c)\times \U(N_f))/ \U(1)$.
This gauging changes the IR scaling dimensions, and also changes the convergence bound for the $S^3$ partition functions.

\end{itemize}

Another highlight of our paper is a formula for the scaling dimension of the quark applicable to any gauge group in the 
large $N_f$ limit, up to the order of $1/N_f^2$ \eqref{Delta}.

The organization of the rest of this paper is given as follows.
In sections \ref{sec.SU}, \ref{sec.USp},  \ref{sec.SO}  
we discuss $\SU(N_c), \USp(N_c)$ and $\SO(N_c)$ SQCD in turn.
In section \ref{sec.group} we briefly comment on group theory aspects
of the scaling dimensions. In section \ref{sec.quiver}
we discuss quiver gauge theories, and in section \ref{sec.M5}
we comments on implications of our results to the theories arising from the compactifications of M5-branes,
and in particular their $1/4$-BPS co-dimension $4$ defects.
The appendices contain several technicalities and review materials.

\section{\texorpdfstring{$\SU(N_c)$ SQCD}{SU(Nc) SQCD}}\label{sec.SU}

Let us begin with $\SU(N_c)$ SQCD with $N_f$ flavors,
and its magnetic dual \cite{Aharony:2013dha}.
Note that in three dimensions even a $\U(1)$ gauge group becomes strongly coupled in the IR, and 
we indeed will find crucial differences from
the case of $\U(N_c)$ SQCD analyzed in \cite{Safdi:2012re}.

\subsection{Dual Pairs}

\paragraph{Electric Theory}
The electric theory has a gauge group (3d $\mathcal{N}=2$ vector multiplet) $\SU(N_c)$,
as well as 
quarks $Q$ in the fundamental representation
and anti-quarks $\tilde{Q}$ in the anti-fundamental representation (these fields are 3d $\mathcal{N}=2$ chiral multiplets).
We do not have a superpotential term: $W_{\rm electric}=0$.
The theory has $N_c-1$ independent monopole operators corresponding to the
Cartan of the gauge group, however
most of them are lifted by the instanton-generated superpotential,
with the exception of a a single unlifted monopole operator which is typically denoted by $Y$ in the literature \cite{Aharony:1997bx} ({\it cf.} Appendix \ref{app.monopole}).
This should be contrasted with the case of a $\U(N_c)$ gauge group,
where we have two unlifted monopole operators $V_{\pm}$ \cite{Aharony:1997bx,deBoer:1997kr}.

The theory has $\SU(N_f)_L \times \SU(N_f)_R \times \U(1)_B \times \U(1)_A\times \U(1)_{\rm R-UV}$ flavor symmetries,
under which the fields $Q, \tilde{Q}, Y$ transform as follows:
\begin{align}
\centering
\bigskip
\begin{tabular}{c||c|ccccc}
 & $\SU(N_c)$ & $\SU(N_f)_L$ & $\SU(N_f)_R$ & $\U(1)_B$ & $\U(1)_A$ & $\U(1)_{\rm R-UV}$ \\
\hline
\hline
$Q$ & $\bm{N_c}$ & $\bm{N_f}$ & $\bm{1}$ & 1 & 1  & 0 \\
$\tilde{Q}$ & $\overline{\bm{N_c}}$ & $\bm{1}$ & $\overline{\bm{N_f}}$ & $-1$ & 1  & 0 \\
\hline
$Y$ & $\bm{1}$ & $\bm{1}$ & $\bm{1}$ & 0 & $-2N_f$  & $2(N_f-N_c+1)$ \\
\end{tabular}
\end{align}
Here the $\U(1)_R$-charge was denoted $\U(1)_{\rm R-UV}$, 
to emphasize that it is one of the many possible $\U(1)$ R-symmetries of the UV theory
and is not the IR $\U(1)$ R-symmetry inside the superconformal algebra.
We listed the $\U(1)_{\rm R-UV}$-charge of the monopole operator $Y$ \cite{Borokhov:2002cg}; 
we will comment more on this later
when we discuss the $S^3$ partition function.

Note also that the theory has no topological $\U(1)_J$ symmetry:
the topological $\U(1)_J$ symmetry is generated by the current $J=*\textrm{Tr}F$, 
however this vanishes since the gauge field is traceless.

\paragraph{Magnetic Theory}
Let us first assume that $N_f>N_c$. 
The electric theory then has a magnetic dual \cite{Aharony:2013dha} (see also \cite{Park:2013wta}).

The gauge group is $\SU(N_f-N_c)\times \U(1)_{\rm diag} \simeq \U(N_f-N_c)$,
and not $\SU(N_f-N_c)$ as one might naively expect.
For notational simplicity 
we define $\tilde{N_c}$ by
\begin{align}
\tilde{N}_c:= N_f-N_c \ .
\end{align}
The theory has dual quark $q$ and anti-quark $\tilde{q}$,
and also $b$ and $\tilde{b}$. 
The meson $M=Q\tilde{Q}$, as well as the monopole operator $Y$
of the electric theory, 
are now fundamental fields in the magnetic theory.
The magnetic theory also has two unlifted monopole operators
$\tilde{X}_{\pm}$. 

The theory also has a superpotential 
\es{Superpotential1}{
W_{\rm magnetic}=M q \tilde{q} + Y b \tilde{b}
+\tilde{X}_{-}+\tilde{X}_{+} \ . 
}
Note that this superpotential 
breaks the 
topological $\U(1)_J$ symmetry, which rotates the fields $\tilde{X}_{\pm}$.

The magnetic theory has the same flavor symmetry as the electric theory,
under which the fields transform as follows:
\begin{align}
\centering
\bigskip
\begin{tabular}{c||cc|ccccc}
 & $\SU(\tilde{N}_c)$ & $\U(1)_{\rm diag}$ &  $\SU(N_f)_L$ & $\SU(N_f)_R$ & $\U(1)_B$ & $\U(1)_A$ & $\U(1)_{\rm R-UV}$ \\
\hline
\hline
$q$ & $\bm{\tilde{N}_c}$ & $\frac{1}{\tilde{N}_c}$ & $\overline{\bm{N_f}}$ & $\bm{1}$ & 0 & $-1$  & $1$ \\
$\tilde{q}$ & $\overline{\bm{\tilde{N}_c}}$ & $-\frac{1}{\tilde{N}_c}$ & $\bm{1}$ & $\bm{N_f}$ & 0 & $-1$  & 1 \\
$b$ & $\bm{1}$ & $-1$ & $\bm{1}$ & $\bm{1}$ & $N_c$ & $N_f$  & $-\tilde{N}_c$ \\
$\tilde{b}$ &  $\bm{1}$ & $1$  & $\bm{1}$ & $\bm{1}$ & $-N_c$ & $N_f$  & $-\tilde{N}_c$ \\
$M$ & $\bm{1}$ & $0$ &$\bm{N_f}$ & $\overline{\bm{N_f}}$ & 0 & $2$  & $0$ \\
$Y$ & $\bm{1}$ &  $0$ &$\bm{1}$ & $\bm{1}$ & 0 & $-2N_f$  & $2(\tilde{N}_c+1)$ \\
\hline
$\tilde{X}_{\pm}$ &  $\bm{1}$ & $0$ & $\bm{1}$ & $\bm{1}$ & 0 & $0$ & 2 \\
\end{tabular}
\label{SU_e_table}
\end{align}
Since $\U(1)_{\rm diag}$ is an Abelian symmetry, there is no canonical normalization of its charges;
the charges above, which differ from those in \cite{Aharony:2013dha} by a factor of $\tilde{N}_c$, are chosen in such a way that it matches with the standard normalization 
when embedded into the $\U(\tilde{N}_c)$ gauge group.

The case of $N_f=N_c$ requires a separate analysis.
In this case the magnetic theory does not have any gauge fields,
and is described by chiral multiplets $Y, M, B, \tilde{B}$ with the 
superpotential
\begin{align}
W=Y(B \tilde{B}-\textrm{det}(M)) \ .
\end{align}
The fields $Y$ and $M$ are the monopole operator and the meson of the electric theory, as before.
The fields $B$ and $\tilde{B}$ are the baryons, which when $N_f>N_c$ are gauge-invariant and 
related to the
$b, \tilde{b}$ of the above-mentioned magnetic theory by the relation
\begin{align}
B=q^{\tilde{N}_c} b \ , \quad \tilde{B}=\tilde{q}^{\tilde{N}_c} \tilde{b} \ .
\end{align}
The charge assignment of the fields $M, Y, B, \tilde{B}$ is 
\begin{align}
\centering
\bigskip
\begin{tabular}{c||ccccc}
 &   $\SU(N_f)_L$ & $\SU(N_f)_R$ & $\U(1)_B$ & $\U(1)_A$ & $\U(1)_{\rm R-UV}$ \\
\hline
\hline
$M$ &$\bm{N_f}$ & $\overline{\bm{N_f}}$ & 0 & $2$  & $0$ \\
$Y$ &  $\bm{1}$ & $\bm{1}$ & 0 & $-2 N_c$  & $2$ \\
$B$ &  $\bm{1}$ & $\bm{1}$ & $N_c$ & $N_c$  & $0$ \\
$\tilde{B}$ &   $\bm{1}$ & $\bm{1}$ & $-N_c$ & $N_c$  & $0$ \\
\end{tabular}
\label{SU_m_table}
\end{align}

The case of $N_f<N_c$ can be derived from the $N_f=N_c$ theory by mass deformation.
For $N_f=N_c-1$ the Coulomb branch smoothly connects with the Higgs branch, giving rise the to constraint
$Y \textrm{det}(M)=1$ \cite{Aharony:1997bx}.
When we have $N_f<N_c-1$, the instanton-generated superpotential 
completely lifts the vacuum moduli space \cite{Affleck:1982as}.
For this reason we will concentrate on the case $N_f\ge N_c$ in the rest of this section.

\subsection{IR Analysis}

As already mentioned in Introduction,
the R-symmetry mentioned above is only one of the many possible R-symmetries in the UV,
and the correct IR R-symmetry inside the superconformal algebra
is a mixture of the UV R-symmetry with global symmetries.
The correct combination is determined by the procedure of $F$-maximization \cite{Jafferis:2010un}.

Since non-Abelian flavor symmetries do not mix with the 
$\U(1)$ R-symmetry, we can parametrize the 
R-symmetry as
\es{IRsymm1}{
R_{\rm IR}=R_{\rm UV}+ a J_A+ b J_B \ ,
}
where $R_{\rm UV}, R_{\rm IR}, J_A, J_B$ are generators of 
$\U(1)_{\rm R-UV}, \U(1)_{\rm R-IR}, \U(1)_A, \U(1)_B$, respectively.

As we will see momentarily $F$-maximization gives $b=0$,
and $b$ does not play crucial roles below.

\paragraph{Unitarity Bound}

The dimensions of the operators $Y, M$ are given by
\es{Dim1}{
&Y: \, \Delta_Y=2(N_f-N_c+1) -2N_f \, a  \ ,\\
&M: \, \Delta_M=2 a \ .
}
The unitary bound $\Delta_{Y,M}\ge \frac{1}{2}$ is given by
\es{Unitarybd1}{
&Y: \, a\le \frac{N_f-N_c+\frac{3}{4}}{N_f} \approx 1-\frac{1}{x}  \ ,\\
&M: \, a\ge \frac{1}{4}  \ ,
}
where here and in the following the symbol $\approx$ will 
denote the Veneziano limit \eqref{Veneziano}.

Note that there are other gauge singlet operators, such as $\tilde{q}q$ and $\tilde{b}b$ in the magnetic theory, 
whose dimension could become
smaller than the threshold value $\frac{1}{2}$. However these are not chiral primary operators,
and hence the constraints from the unitarity bound does not necessarily apply. For example,
the operator $\tilde{q}q$ is trivial in the chiral ring thanks to the F-term relation for the field $M$,
and hence is not a chiral primary. The same applies to the operator $\tilde{b}b$.

\paragraph{$F$-maximization}

The $S^3$ partition function \cite{Kapustin:2009kz,Jafferis:2010un,Hama:2010av}
of the electric/magnetic theories can be written down straightforwardly following the 
matter content given above (see Appendix \ref{app.S3}).
For the electric theory we have
\es{ElectricPF1}{
\begin{split}
Z_{\rm electric}&=
\frac{1}{N_c!}\int \prod_{i=1}^{N_c} d\sigma_i   \, 
\delta\left(\sum_{i=1}^{N_c} \sigma_i\right)
\overbrace{\prod_{1\le i<j\le N_c} \sinh^2[\pi(\sigma_i-\sigma_j)]}^{\text{measure}} \\
&\quad \times
\overbrace{\prod_{i=1}^{N_c} \exp\left[
  N_f \, l(1-a\pm b \pm i \sigma_i ) 
  \right]}^{Q, \, \tilde{Q}} \ .
\end{split}
}
The integral is over the Cartan of the $\SU(N_c)$ gauge group,
and the integrand represents the 1-loop determinants for $\mathcal{N}=2$ 
vector and chiral multiplets.
Here and in the following we use the shorthanded notation
that $\pm$ inside the expression means the sum of the corresponding two expressions.
For example, 
\begin{align}
l(1-a\pm b \pm i \sigma_i ) :=l(1-a+b + i \sigma_i )+l(1-a- b - i \sigma_i ) \ .
\end{align}

For the magnetic theory, let us first consider the case $N_f>N_c$. We then have
\es{MagneticPF0}{
Z_{\rm magnetic}& =
\frac{1}{ \tilde{N}_c!}\exp\left[ \overbrace{N_f^2 \, l(1-2a)}^{M} +\overbrace{ l\left(1+2N_f \, a-2(\tilde{N}_c+1) \right) }^{Y}\right]  \\
&\times
\int d\sigma \, 
\int \prod_{i=1}^{\tilde{N}_c} d\sigma_i \,\, \delta\left(\sum_{i=1}^{N_c} \sigma_i\right) \,
\overbrace{\prod_{1\le i<j\le \tilde{N}_c} \sinh^2[\pi(\sigma_i-\sigma_j)] }^{\rm measure} \\
&\times \overbrace{ \prod_{i=1}^{\tilde{N}_c} 
\exp\left[  
    N_f \,  l\left(a\pm  i \sigma_i \pm i \frac{1}{\tilde{N}_c}\sigma\right)
\right] }^{q,  \, \tilde{q}}
\times \overbrace{\exp\left[  
     l\left(1+\tilde{N}_c-N_f \, a \pm N_c\, b \mp i  \sigma \right)
    \right] 
    }^{ b, \, \tilde{b}}
 \ ,    
}
where $\sigma_i$ ($\sigma$) parametrizes the Cartan of $\SU(\tilde{N}_c)$ ($\U(1)_{\rm diag}$).
The $\sigma$-dependence inside the integrand can be 
eliminated by the shift $\sigma_i\to \sigma_i-\frac{1}{N_c} \sigma$, 
after which the delta function constraint becomes $\sigma=\sum_{i=1}^{\tilde{N}_c}\sigma_i$, i.e.\
$\sigma$ is the diagonal part of the $\U(\tilde{N}_c)$ gauge group. 
After a trivial delta-function integral over $\sigma$ we obtain
\es{MagneticPF1}{
\begin{split}
Z_{\rm magnetic}& =
\frac{1}{ (N_f - N_c)!}\exp\left[ N_f^2 \, l(1-2a) + l\left(1+2N_f \, a-2(N_f-N_c+1) \right) \right]  \\
&\times 
\int \prod_{i=1}^{\tilde{N}_c} d\sigma_i \,
\prod_{1\le i<j\le N_f-N_c} \sinh^2[\pi(\sigma_i-\sigma_j)]  \\
& \times \prod_{i=1}^{N_f-N_c} 
\exp\left[  
    N_f \,  l\left(a\pm  i \sigma_i \right)
\right] 
\times \exp\left[  
     l\left(1+\tilde{N}_c-N_f \, a \pm N_c\, b \mp i  \sum_{i=1}^{\tilde{N}_c} \sigma_i \right)
    \right] \ .
    \end{split} 
}

The case of  $N_f=N_c$ is much simpler thanks to the absence of the gauge group
in the magnetic theory. We have
\begin{align}
Z^{N_f=N_c}_{\rm magnetic}=\exp\left[\overbrace{N_f^2\, l(1-2a)}^M+\overbrace{ l(1-2+2N_c a )}^{Y}+\overbrace{l(1-N_c( a\pm b))}^{B, \, \tilde{B}}\right] \ .
\end{align}
Note that this expression can also be obtained by
formally setting $N_f=N_c$ in \eqref{MagneticPF1}.

\begin{figure}[htbp]
\begin{center}
\includegraphics[scale=2.0]{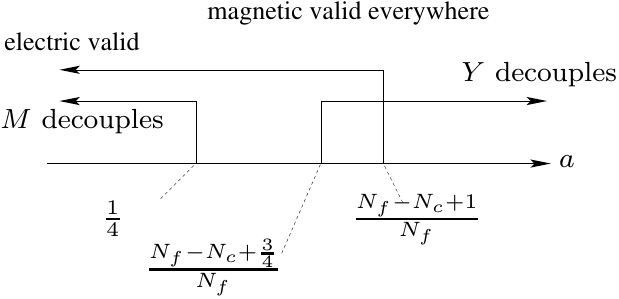}
\caption{The unitarity bound and the convergence bound for 3d $\mathcal{N}=2$ $\SU(N_c)$ SQCD with $N_f$ flavors with $N_f> N_c$, plotted in terms of the mixing parameter $a$ (see \eqref{IRsymm1}, with $b=0$). The correct IR value of $a$ should be determined from $F$-maximization.
}
\label{fig.SU}
\end{center}
\end{figure}

\paragraph{Convergence}

We have written down the expressions for the $S^3$ partition function,
however they are in general only formal integral expressions and 
are actually not convergent.

We can analyze the convergence condition of the partition function
by sending one of the $\sigma_i$'s to infinity
(for the electric theory, we need to send one to infinity and another to minus infinity,
for the consistency with the traceless constraint $\sum_i \sigma_i=0$), 
while keeping other $\sigma_i$'s finite. We can evaluate the leading behavior of the integrand from the 
asymptotic expansions (see \eqref{Expansion1} in Appendix \ref{app.S3}), and
we see that the convergence bound of the partition function is
\es{Convergebd1}{
&\textrm{electric: } \quad a <  \frac{N_f-N_c+1}{N_f} \approx 1-\frac{1}{x}  \ ,  \\
&\textrm{magnetic: converges for any values of $a$}  \ .
}
Note that for numerical computations the practical convergence bound is slightly stronger than this, since as we approach the convergence bound the 
computational time becomes increasingly large.

It turns out that these conditions are the same as the condition that the 
dimensions of the monopole operators ($Y$ for the electric theory, $\tilde{X}_{\pm}$
for the magnetic theory) are non-negative:
\es{DimYX}{
&Y: \, \Delta_Y=2(N_f-N_c+1) -2N_f \, a \ge 0  \ ,\\
&\tilde{X}_{\pm}: \, \Delta_{\tilde{X}_{\pm}}=2 \ge 0 \ .
}

That we obtain the same conditions from two different considerations is not a coincidence,
and we will encounter the same phenomena in later sections. In fact, we can think of this as a convenient way to 
obtain the R-charge/conformal dimension of monopole operators.

When we analyze the convergence of the partition function, we go to infinity in the Coulomb branch
in the direction of the Cartan 
 corresponding to 
a monopole operator $V$,
for example $\sigma_1=-\sigma_N \to \infty, \sigma_{j\ne 1,N}=0$ in the magnetic theory ({\it cf.} Appendix \ref{app.monopole}).
Since Coulomb branch parameter is a dynamical version of the real mass parameter, this has the effect of making 
the fields massive. We can integrate out the these massive modes, except that we then could have 
induced Chern-Simons term with level $k_{\rm eff}$ and induced FI parameter $\zeta_{\rm eff}$.
In the theories discussed in this paper, we have $k_{\rm eft}=0$ however $\zeta_{\rm eff}\ne 0$, 
leaving to the expression
\begin{align}
Z\sim \int d\sigma_1 \,\,  e^{-2\pi \zeta_{\rm eff} \sigma_1}   \ ,
\end{align}
and the dimension (or equivalently the $\U(1)_R$-charge) of the monopole operator $V$, whose real part is $e^{-2\pi \sigma_1}$,
can be identified with $\zeta_{\rm eff}$:
\begin{align}
\label{zetaR}
\zeta_{\rm eff}=\Delta_V \ . 
\end{align}
In our example, the partition function gives
\begin{align}
&\textrm{electric:  } \quad\zeta_{\rm eff}=-\overbrace{(N_c-1)}^{\text{measure}}+\overbrace{N_f (1-a)}^{Q, \, \tilde{Q}} 
=N_f-N_c+1-N_f \, a \ , \\
&\textrm{magnetic:  }\quad \zeta_{\rm eff}=-\overbrace{(\tilde{N}_c-1)}^{\text{measure}}+\overbrace{N_f a}^{q, \, \tilde{q}}
+\overbrace{1+\tilde{N}_c-N_f\, a}^{b, \, \tilde{b}} =2 \ , 
\end{align}
and their positivity conditions indeed match with \eqref{DimYX}.

As a side remark,
this also explains clearly that the convergence condition is weaker than the unitarity constraint:
for a monopole operator $V$ 
the former requires $\Delta_V\ge 0$, while the latter requires $\Delta_V\ge \frac{1}{2}$.
 
\paragraph{Duality as Equality}

In the literature, the duality between two 3d $\mathcal{N}=2$ theories is often 
translated to an equivalence of the 
$S^3$ partition functions:
\begin{align}
Z_{\rm electric} =  Z_{\rm magnetic} \ .
\label{Zduality}
\end{align}
Such equivalences have been verified in \cite{Dolan:2011rp,Willett:2011gp,Bashkirov:2011vy,Benini:2011mf,Aharony:2013dha}.\footnote{In Appendix \ref{app.UtoSU} we derive such an identity fro the $\SU(N_c)$ theory
from the corresponding identity for the $\U(N_c)$ theory.}

There are subtleties to the identity \eqref{Zduality}, however.
In fact, as we have already seen in \eqref{Convergebd1},
we find that as a function of the parameter $a$ 
the right hand side always converges, whereas
the left hand side converges only when $a$ is small enough,
invalidating \eqref{Zduality}.

We can see this problem more sharply 
for the $N_f=N_c$ theory.
The magnetic partition function has poles at
$2 N_c a=\bZ\backslash \{1 \}$, as follows from the 
definition of the function $l(z)$ (see Appendix \ref{app.S3}).
The electric partition function, however, does not show any singular behavior at these points.

This is not in contradiction with the existing results in the literature.
In the analysis above we assumed that $a$ is a real parameter, however 
in the literature $a$ takes values in the complex plane, 
where the imaginary part of $a$ plays the role of the real mass parameter
for the $\U(1)_A$ flavor symmetry.
The $S^3$ partition function is known to be a holomorphic function
of this complexified parameter \cite{Jafferis:2010un,Festuccia:2011ws},
and we can then regard the both sides of \eqref{Zduality} as 
complex functions of $a$, establish the identities in the regions where the real part of $a$ is small, and then 
analytically continue into the whole complex plane.
This is what is usually meant by the identity \eqref{Zduality}.

However, we do not wish to turn on imaginary parts of $a$
for the purpose of this paper. When we turn on the real mass parameter
for the $\U(1)_A$ symmetry, the quarks $Q, \tilde{Q}$ gets a mass and hence can be integrated out in the deep IR,
thereby dramatically changing the IR behavior of the theory.
We need to keep $a$ real, for the numerical analysis of the $F$-maximization below.

This means that we need be careful in interpreting the equality
\eqref{Zduality}, at least for the purpose of $F$-maximization---When only one of the two sides converge, we should use that convergent partition function to determine the IR conformal dimensions,
whereas if the both sides converge they should give the same value of the $F$-function (possibly up to an overall constant independent of the parameter $a$) and they both give the same IR conformal dimensions.\footnote{It is possible to render the expression convergent by deforming the integration contour at infinity. However, the meaning of such a deformation, and the connection with the $F$-theorem, is not clear.}

In the case of the $\SU(N_c)$ SQCD discussed here,
the result \eqref{Convergebd1} shows that 
magnetic partition function is convergent for all the values of the parameter $a$.
This is in sharp contrast with the case of the $\U(N_c)$ SQCD discussed in \cite{Safdi:2012re},
where region for the convergent magnetic partition function was
complementary to that of the electric partition function;
the overlapping region exists only for fine-tuned values of $N_c$ and $N_f$,
and vanishes in the Veneziano limit.

\paragraph{$F$-maximization}
We can determine the values of $a$ and $b$ by maximization
of the free energy $F$, related to the $S^3$ the partition function
\eqref{ElectricPF1} and \eqref{MagneticPF1} (which are identical thanks to the duality, modulo the issues just mentioned)
by the relation
\begin{align}
F:=-\log | Z_{S^3} | \ .
\end{align}
Maximization with respect to $b$ straightforwardly gives $b=0$.
We can then numerically search for the maximal value of $F$
with respect to $a$. Note that it is crucial for our numerical analysis that $F$ takes a maximal value, 
not just an extremal value. In fact, in many of our examples the function $F$ has
more than one local maximums.

We have carried out explicit numerical integration of the matrix integral, and obtained the critical value of $a$
after $F$-maximization, for some sample values of small $N_c$ and $N_f\ge N_c$,
as shown in Table \ref{SUtable}.

\begin{table}
\begin{center}
\begin{tabular}{| c | ccccc |}
 \hline
    & $N_f = 2$ & $N_f = 3$ & $N_f = 4$ & $N_f = 5$ & $N_f = 6$  \\
   \cline{1-6} $N_c = 2$  & \spc{$.2862$  \\$(5.3743)$} & \spc{$.3687$ \\ $(7.6517)$ } &  \spc{$.4064$ \\ $(9.6090)$}  & \spc{$.4277$\\ $(11.4143)$} &  \spc{$.4412$ \\ $(13.1306)$} \\
  $N_c = 3$ &  - & \fcolorbox{red}{white}{\spc{$.2222$ $[M]$ \\$(3.9915)$}} & \spc{$.322$ \\ $(14.0933)$ } & \spc{$.3632$ \\ $(17.4956)$} &  \spc{$.3898$ \\$(20.5698)$ } \\
  $N_c = 4$ & - & - & \fcolorbox{red}{white}{\spc{$\frac{1}{8}$ $[MY, B]$ \\$(6.5849)$}}  & \spc{$.2912$ \\ $(22.4819)$} & \spc{$.3353$ \\ $(27.5200)$} \\
 $N_c = 5$ & - & - & - & \fcolorbox{red}{white}{\spc{$\frac{1}{10}$$[MY, B]$ \\$(9.7041)$} } & \fcolorbox{blue}{white}{\spc{$.2693$ \\$(16.9044)$}}\\
  $N_c = 6$ & - & - & - & - & \fcolorbox{red}{white}{\spc{$\frac{1}{12}$ $[Y, M, B]$ \\$(13.5164)$}} \\
  \hline
\end{tabular}
\end{center}
\caption{The scaling dimension $\Delta_Q$ of the flavor multiplets (above) and the maximal value of $F$-function (below), at the conformal fixed points for a few small values of $N_f$ and $N_c$ in 3d ${\cal N} = 2$ $\SU(N_c)$ SQCD with $N_f$ flavors.
We have computed this from the electric theory, except for the diagonal entries and the blue-colored entries
where we used a simpler magnetic theory for more efficient numerical evaluation.
For $N_c=N_f=3, 4, 5, 6$ either one or two operators hit the unitarity bound, and consequently we need to decouple them
and repeat the $F$-maximization with the modified $F$-function \eqref{FY}, until the procedure terminates.
For $N_c=N_f=4, 5, 6$ we find a sequence of decoupling of operators, leaving to a free IR theory eventually---for example, for 
$N_c=N_f=4$ operators $M$ and $Y$ decouple first, and the baryon $B$ becomes free after the second $F$-maximization, Similarly, for $N_c=N_f=6$ we find that first $Y$ decouples, then $M$, and finally $B$ becomes free.
Such a decoupling pattern is shown inside the bracket in the red box.
Note that the value of the scaling dimension $\Delta_Q$ shown here is the value after all the possible decoupling effects are taken into account, and not the value after the first $F$-maximization.}
\label{SUtable} 
\end{table}

For the values $N_f=N_c=3, 4, 5, 6$ (in entries in red boxes in Table \ref{SUtable})
we find that after performing the first $F$-maximization that 
the unitarity bound \eqref{Unitarybd1} is violated for some operators.
We interpret this to mean that we need to decouple corresponding operators.
The details of the decoupling varies for different values of
$N_f$ and $N_c$, as shown in the diagonal entries of 
Table \ref{SUtable}. 

After decoupling an operator, we need to again do $F$-maximization with the 
modified $F$-function, which are for example given by 
\begin{align}\label{FY}
\begin{split}
F_{\rm magnetic}^\textrm{$M$ decoupling}&=N_c^2 \, l\left(\frac{1}{2}\right)+l\left(2N_c a-1\right) +2\,  l(1-N_c a) \ , \\
F_{\rm magnetic}^\textrm{$Y$ decoupling}&=N_c^2 \, l(1-2a)+l\left(\frac{1}{2}\right) +2\,  l(1-N_c a) \ , \\
F_{\rm magnetic}^\textrm{$MY$ decoupling}&=N_c^2 \, l\left(\frac{1}{2}\right)+l\left(\frac{1}{2}\right) +2\,  l(1-N_c a) \ ,
\end{split}
\end{align}
depending on whether only $M$, only $Y$ or both $M$ and $Y$ decoupling. 
For the choice of $N_f=N_c=4, 5, 6$ we find after the second (and third for $N_f=N_c=6$) $F$-maximization
that we need to decouple further operators,
and consequently find that all the operators $M, Y, B$ become free, leaving to a free IR theory.

Something interesting happens for $N_f =N_c \ge 6$. After the first $F$-maximization, we find that the monopole operator $Y$ decouples. After decoupling $Y$, we find that the modified $F$-function apparently has no maximum.
We propose to interpret this as a signal for the decoupling of the baryon $B$. After yet another $F$-maximization 
we find that the meson $M$ also becomes free, leading to the critical value $a=1/4$ and the trivial IR fixed point.
One consistency check of this proposal is that the critical value $a=1/4$ is consistent with the 
analysis of the Veneziano limit shown below in Figure \ref{fig:CrackSU}.

Note also that the
value of the $F$ at the critical value decreases as 
we decrease the value of $N_f$. This is consistent with the $F$-theorem \cite{Myers:2010xs,Jafferis:2011zi,Klebanov:2011gs,Casini:2012ei},
since we can give a mass to one of the flavors, thereby
 reducing the number of flavors by one.
It is probably worth pointing out that for a fixed flavor number $N_f$ the value of the free energy
$F$ could decrease as we increase $N_c$.

\bigskip

In Table \ref{SUtable} the monopole operator decoupling happens only in the diagonal $N_f=N_c$.
However this is an artifact of the choice of small $N_c, N_f$ values.
The constraint from the unitarity bound \eqref{Dim1}
becomes stronger as we increase the value of $N_c, N_f$, and therefore we expect to find more and more examples of
$(N_c, N_f)$ with monopole decoupling.

To analyze monopole decoupling in the Veneziano limit,
we adopt techniques of \cite[Appendix A.3]{Safdi:2012re} (see also \cite{Klebanov:2011td}),
which gives the scaling dimension $\Delta_Q$ of the 
quarks/anti-quarks to be 
\es{DeltaSU}{
\Delta_Q(N_c, N_f) &= \frac{1}{2} - \frac{2(N_c - \frac{1}{N_c})}{\pi^2 N_f} + \frac{(24N_c^2 - 48 + \frac{24}{N_c^2}) - \pi^2 (\frac{10N_c^2}{3} -\frac{16}{3} +\frac{2}{N_c^2}) }{ \pi^4 N_f^2}+ \mathcal{O}\left(\frac{1}{N_f^3}\right) \,,
}
which reduces to
\es{DeltaSUlimit}{
\Delta_Q(x) = \frac{1}{2} - \frac{2 }{\pi^2 x} + \frac{8(36 - 5\pi^2) }{ 12\pi^4 x^2}  + \mathcal{O}\left(\frac{1 }{ x^3}\right) \,. 
}

The combined plot of the numerical data points as well as the 
large $x$ expansion of \eqref{DeltaSUlimit} is shown in 
Figure \ref{fig:CrackSU}.
We find that $\Delta_Q(x)$ hits the convergence bound at the critical value $x_c \approx 1.46$.

\begin{figure}[htb]   
\leavevmode
\begin{center}$
\begin{array}{cc}
\scalebox{.65}{\includegraphics{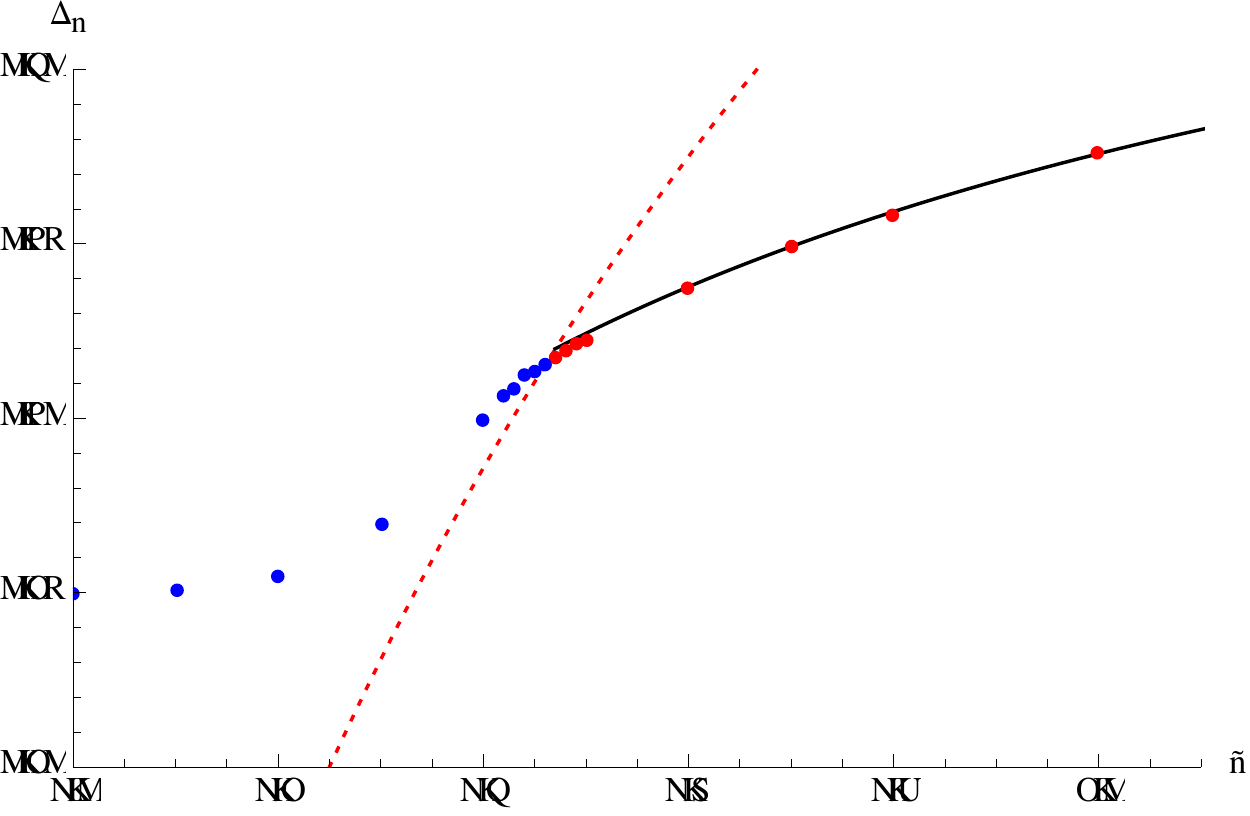}} & \scalebox{.65}{\includegraphics{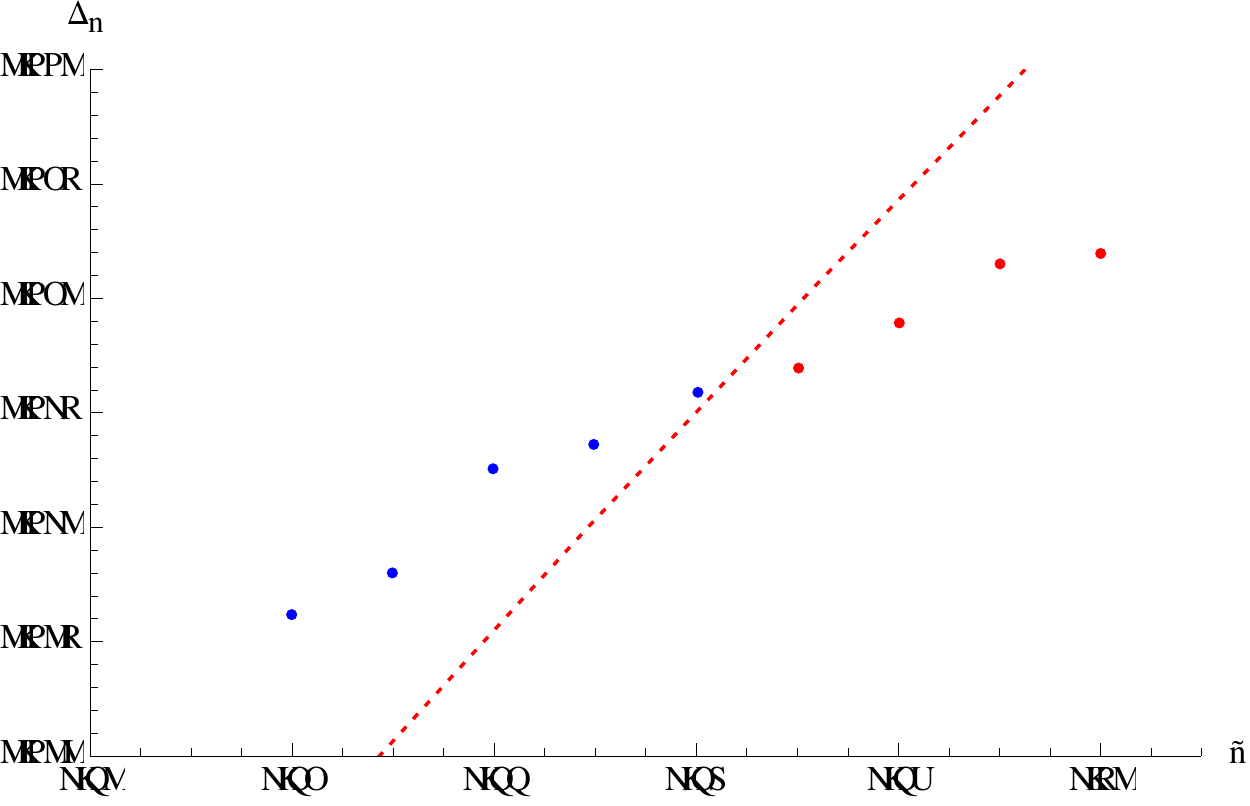}}
\end{array}$
\end{center}
\caption{$\Delta_Q$ as a function of $x=N_f/N_c$ in the Veneziano limit. Points were computed by extrapolating small $N_c$ numerical results. Dotted line is the unitarity bound~\eqref{Unitarybd1}. We find that $\Delta_Q(x)$ hits the unitarity bound at the critical value $x_c \approx 1.46$. The black curve at large values of $x$ is the analytical approximation~\eqref{DeltaSUlimit}. In the region right to the red curve, we use electric theory, while in the left region we use magnetic theory with monopole $Y$ decoupled when needed. The right plot is a zoomed-in version of the left one around the critical value $x_c$.}
\label{fig:CrackSU}
\end{figure}

The analysis in this subsection is partly case-by-case, and 
it would be interesting to find more uniform patterns in their IR behaviors.
A related question is to find a concrete UV Lagrangian description of the theory 
after decoupling of monopole operators, perhaps along the lines of \cite{Yaakov:2013fza}.

\section{\texorpdfstring{$\USp(2N_c)$ SQCD}{USp(2Nc) SQCD}}\label{sec.USp}

Let us next consider the $\USp(2N_c)$ theory.\footnote{Readers not interested in $\USp$ or $\SO$ gauge groups can 
proceed directly to the discussion of quiver gauge theories in section \ref{sec.quiver}.}
We find that the structure here is similar to the case of the $\U(N_c)$ theory.
In particular, we find a small window where
the electric and magnetic descriptions hold simultaneously,
which window shrinks in the Veneziano limit.

\subsection{Dual Pairs}
\paragraph{Electric Theory}
The electric theory is given by
quarks $Q$, and comes with a monopole operator $Y$.
We do not have a superpotential term: $W_{\rm electric}=0$.
The theory has $\SU(2 N_f) \times \U(1)_A\times \U(1)_{\rm R-UV}$ flavor symmetry,
under which the quark $Q$ and the monopole operator $Y$ transform as follows:
\begin{align}
\begin{tabular}{c||c|ccc}
 & $\USp(2N_c)$ & $\SU(2 N_f)$ & $\U(1)_A$ & $\U(1)_{R-UV}$  \\
\hline
\hline
$Q$ & $\bm{2 N_c}$ & $\bm{2 N_f}$ & $1$ & 0  \\
\hline
$Y$ & $\bm{1}$ & $\bm{1}$ & $-2 N_f$ & $2(N_f-N_c)$ \\
\end{tabular}
\end{align}

\paragraph{Magnetic Theory}
For $N_f > N_c+1$,
the dual magnetic theory has $\USp(2N_f - 2N_c -2)$ gauge symmetry with $2N_f$ chiral multiplets, dual quark $q_i$, and additional single chiral multiplets $M$ and $Y$\cite{Aharony:1997gp}. 
Coulomb branch of this magnetic theory is parametrized by the monopole operator $\tilde{Y}$.

The charge assignment is given by
\es{USpDual}{
\begin{tabular}{c||c|ccc}
 & $\USp(2N_f-2N_c-2)$ & $\SU(2 N_f)$ & $\U(1)_A$ & $\U(1)_{R-UV}$  \\
\hline
\hline
$q$ & $\bm{2 N_c}$ & $\bm{2N_f}$ & $-1$ & 1  \\
$M$ & $\bm{1}$ & $\bm{N_f(2N_f-1)}$ & $2$ & 0 \\
$Y$ & $\bm{1}$ & $\bm{1}$ & $-2 N_f$ & $2(N_f-N_c)$ \\
\hline
$\tilde{Y}$ & $\bm{1}$ & $\bm{1}$ & $2 N_f$ & $-2(N_f-N_c-1)$ \\
\end{tabular}
}

The theory also has a superpotential 
\begin{align}
W_{\rm magnetic}=M q q + Y  \tilde{Y} \ . 
\end{align}

For $N_f=N_c+1$, we expect that the magnetic theory is trivial. We propose that the 
magnetic theory in this case is described by $Y$ and $M$, with the superpotential
\begin{align}
W = Y\, \textrm{Pf}\, (M) \;.
\end{align}
We can verify that this theory is consistent with the charge assignment,
which is given as
\es{USpDual_2}{
\begin{tabular}{c||c|ccc}
 &  $\SU(2 N_f)$ & $\U(1)_A$ & $\U(1)_{R-UV}$  \\
\hline
\hline
$M$  & $\bm{N_f(2N_f-1)}$ & $2$ & 0 \\
$Y$ & $\bm{1}$ & $-2 N_f$ & $2$ 
\end{tabular}
}

We have a deformed moduli space $Y\, \textrm{Pf}\, (M)=1$  for $N_f=N_c$ \cite{Karch:1997ux},
and the supersymmetry is broken for $N_f<N_c$. We therefore concentrate on the case $N_f> N_c$ below.

\subsection{IR Analysis}

Let us parametrize the IR R-symmetry by
\es{IRsymm1-USp}{
R_{\rm IR}=R_{\rm UV}+ a J_A \ ,
}
where the notation is the same as in the previous section.

\paragraph{Unitarity Bound}

The dimensions of the operators $Y$ and $M$ are
\es{Dim1-USp_new}{
\Delta_{Y}=2(N_f-N_c)-2 N_f a  \ ,\qquad
\Delta_M = 2a \ .
}
The unitary bound is satisfied if
\es{Dim1-USp}{
&Y: \, a\le \frac{N_f-N_c-\frac{1}{4}}{N_f} \approx 1-\frac{1}{x}  \ ,\\
&M: \, a\ge \frac{1}{4} \ .
}
\paragraph{Partition Function}
The $S^3$ partition function of the electric theory is given by (see \eqref{Cartan} and \eqref{roots})
\es{ElectricPF-USp}{
\begin{split}
Z_{\rm electric}&=
{1 \over 2^{N_c}N_c!}\int \prod_{i=1}^{N_c} d\sigma_i   \, 
\prod_{1\le i<j\le N_c} [2\sinh[\pi(\sigma_i+\sigma_j)]2\sinh[\pi(\sigma_i-\sigma_j)]]^2 \\
&\quad \times
\prod_{i=1}^{N_c} [2\sinh(2\pi\sigma_i)]^2
\overbrace{
 \exp\left[
  2N_f \, l(1-a \pm i \sigma_i ) 
\right]}^{Q} 
\ .
\end{split}
}
For the magnetic theory, we have
\es{MagneticPF-USp}{
\begin{split} 
Z_{\rm magnetic}&=
\frac{1}{2^{N_f-N_c-1}(N_f-N_c-1)!} \\
& \times \exp\left[\overbrace{l(1-2(N_f-N_c)+2N_f a)}^Y+\overbrace{ N_f (2 N_f-1) l(1-2a)}^M\right]\\
&\times \int \prod_{i=1}^{N_f-N_c-1} d\sigma_i  
\prod_{1\le i<j\le N_f-N_c-1} [2\sinh[\pi(\sigma_i+\sigma_j)]2\sinh[\pi(\sigma_i-\sigma_j)]]^2 \\
&\times
\prod_{i=1}^{N_f-N_c-1} [2\sinh(2\pi\sigma_i)]^2 
\overbrace{\exp\left[
  2N_f \, l(a \pm i \sigma_i ) 
\right]}^{q} 
\end{split}
} for $N_f> N_c+1$ and 
\es{MagneticPF-USp_2}{
Z_{\rm magnetic}&=
\exp\left[\overbrace{l(-1+2N_f a)}^Y+\overbrace{N_f (2 N_f-1)  l(1-2a)}^M\right]
}
for $N_f=N_c+1$.

Using again the expansion~\eqref{Expansion1}, we determine the convergence 
bound to be
\es{Convergebd-USp}{
&\textrm{electric: } \quad a< \frac{N_f-N_c}{N_f} \approx 1-\frac{1}{x}  \ ,  \\
&\textrm{magnetic: } \quad a> 
\frac{N_f-N_c-1}{N_f}
\approx 1-\frac{1}{x}  \ .
}
The width of intersection of these two regions shrinks to zero in the 
Veneziano limit.

\begin{figure}[htbp]
\begin{center}
  \includegraphics[scale=1.0]{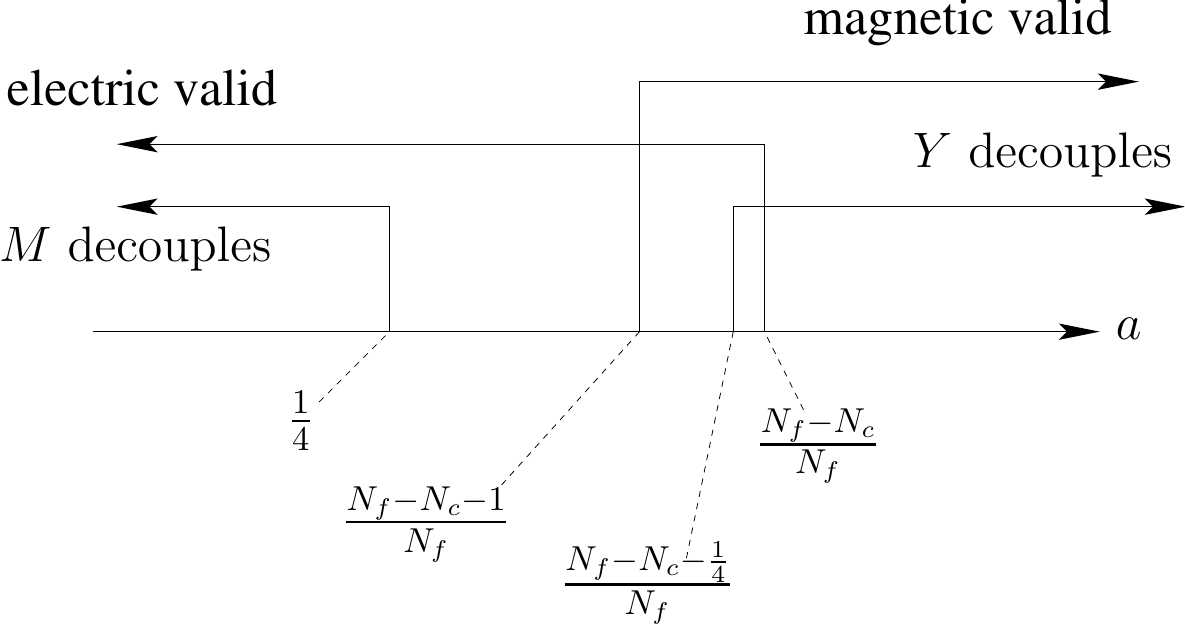}
  \caption{
  The unitarity bound and the convergence bound for 3d $\mathcal{N}=2$ $\USp(2 N_c)$ SQCD with $N_f$ flavors with $N_f> N_c+1$, plotted in terms of the mixing parameter $a$ (see \eqref{IRsymm1-USp}). The correct IR value of $a$ should be determined from $F$-maximization.
The structure here is very similar to the $\U(N_c)$ SQCD case discussed in Appendix \ref{app.UN} (Figure \ref{fig.U}).}
  \label{fig.USp}
  \end{center}
  \end{figure}

\paragraph{$F$-maximization}

We can again numerically maximize the $F$-function for small values of $N_c$ and $N_f$.
For this purpose it is sometimes useful to use the trick explained in 
Appendix \ref{app.numerical} (the same trick could be applied to the $\SO(N_c)$ theory discussed in the next section).
The results of the numerical computation is summarized in Table \ref{USptable}.

\begin{table}
\begin{center}
\begin{tabular}{| c | ccccc |}
 \hline
    & $2N_f = 4$ & $2N_f = 6$ & $2N_f = 8$ & $2N_f = 10$ & $2N_f = 12$  \\
   \cline{1-6} $2N_c = 2$  & \spc{$.2861$ \\ $(2.1502)$ } & \spc{$.3687$ \\ $(4.4275)$ } &  \spc{$.4064$ \\ $(6.3848)$ }  & \spc{$.4277$ \\ $(8.1901)$} &  \spc{$.4412$ \\ $(9.9064)$} \\
  $2N_c = 4$ &  - & \fcolorbox{red}{white}{\spc{$\frac{1}{4}$ $[Y]$\\$(5.5452)$} } & \spc{$.3305$ \\ $(10.8766)$ } & \spc{$.3711$ \\$(15.2865)$ } &  \spc{$.3962$ \\ $(19.2927)$ } \\
  $2N_c = 6$ & - & - & \fcolorbox{red}{white}{\spc{$\frac{1}{4}$ $[Y]$\\$(10.0506)$}} & \spc{$.3038$ \\ $(18.8447)$} & \spc{$.3456$ \\ $(26.1131)$} \\
 $2N_c = 8$ & - & - & - & \fcolorbox{red}{white}{\spc{$\frac{1}{4}$ $[Y]$\\ $(15.9423)$}} & \fcolorbox{blue}{white}{\spc{$.2822$ \\ $(27.9908)$}}\\
  $2N_c = 10$ & - & - & - & - & \fcolorbox{red}{white}{\spc{$\frac{1}{4}$ $[Y]$\\$(23.2204)$}} \\
  \hline
\end{tabular}
\end{center}
\caption{The scaling dimension $\Delta_Q$ of the flavor multiplets (above) and the maximal value of $F$-function (below), at the conformal fixed points for a few small values of $N_f$ and $N_c$ in 3d ${\cal N} = 2$ $\USp(2N_c)$ SQCD with $N_f$ flavors.
For the red boxes in the diagonal (i.e. $2N_f=2(N_c+1)$) entries we find after the first $F$-maximization
that we need to decouple the monopole operator $Y$.
In most of the entries we used the electric partition functions, except in the red-colored (along the diagonal)
and blue-colored (at $2N_c=8, 2N_f=12$) entries
we used magnetic partition functions, since
the magnetic description is more suitable for numerical computations.
}
\label{USptable}
\end{table}

For $N_f =N_c> 2$, we always see that the monopole $Y$ always saturates the unitarity bound, thus we set its scaling dimension at $\frac{1}{2}$. This modified magnetic theory forces $a = \frac{1}{4}$.
Inside the table $Y$ is the only operator which decouples in the IR, and in this sense the structure here is much simpler than 
that of the $\SU(N_c)$ SQCD discussed in previous section.

We can again check the consistency with the $F$-theorem by decreasing the values of $N_f$ for a fixed $N_c$.

We can also obtain analytic expressions of the scaling dimensions in the large $N_f$ limit, 
by the techniques of \cite[Appendix A.3]{Safdi:2012re}.
This gives the scaling dimension of the electric quarks to be
\es{DeltaUSp}{
\Delta_Q(N_c, N_f) &= \frac{1}{2} - {4 N_c + 2 \over 2\pi^2 N_f} + {3(96N_c^2 + 96N_c + 24) - \pi^2 (40N_c^2 +48N_c +14) \over 12\pi^4 N_f^2}+ \mathcal{O}\left({1 \over N_f^3}\right) \,.
}
In the Veneziano limit, this reduces into
\es{DeltaUSplimit}{
\Delta_Q(x) = {1 \over 2} - {2 \over \pi^2 x} + {8(36 - 5\pi^2) \over 12\pi^4 x^2}  + \mathcal{O}\left({1 \over x^3}\right) \,. 
}
On the other hand, when $x$ is close to 1, $(x-1)$ expansion \cite[Appendix A.3]{Safdi:2012re} makes sense and we get 
another expansion of 
$\Delta_Q(x)$:
\es{DeltaUSp_mag}{
\Delta_Q(x) = {1 \over 4} + {1 \over 4\pi}(x-1) + {4 - 18\pi + 5\pi^2 \over 4(2-\pi)\pi^2}(x-1)^2 + \mathcal{O}( (x-1)^3) \ .
}

\begin{figure}[h]   
\leavevmode
\begin{center}
\scalebox{.78}{\includegraphics{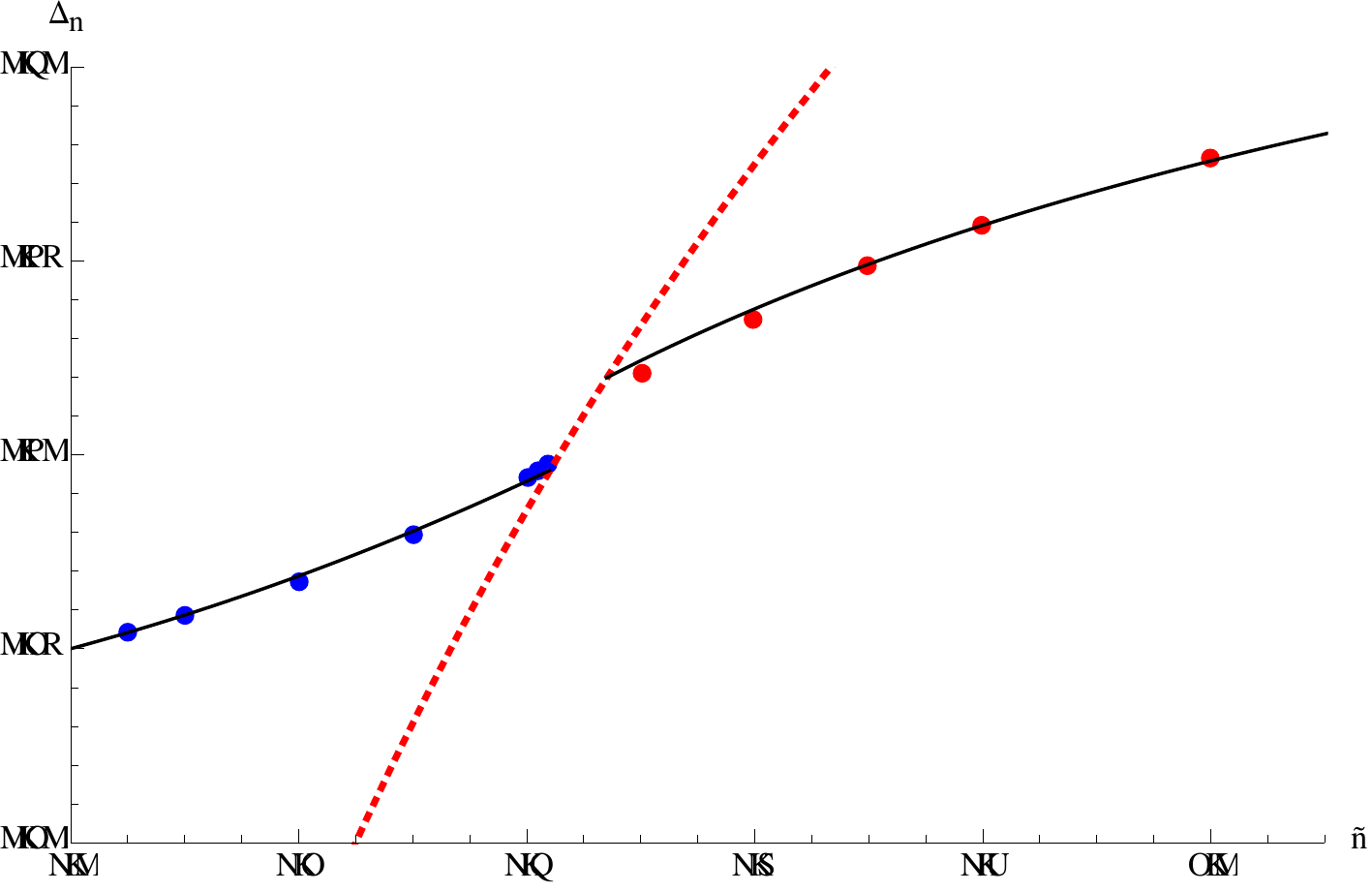}}
\end{center}
\caption{$\Delta_Q$ as a function of $x=N_f/N_c$ in the Veneziano limit. Points were computed by extrapolating small $N_c$ numerical results. Dotted line is the convergence bound~\eqref{Convergebd-USp}. We find that $\Delta_Q(x)$ hits the convergence bound at the critical value $x_c \approx 1.43$. The black curves at large and small values of $x$ are the analytical approximation~\eqref{DeltaUSplimit} and~\eqref{DeltaUSp_mag}, respectively. In the region right to the red curve, we use electric theory (coloured in red), while in the left region we use magnetic theory with monopole $Y$ decoupled (colored blue).}
\label{fig: CrackUSp}
\end{figure}

In Figure \ref{fig: CrackUSp}
we have plotted these result in combination from the numerical data points coming from the 
several explicit integrations for small values of $N_c$ and $N_f$.
We find good agreement between numerical and analytical results, and the critical value for $x$ is 
given by $x_c\approx 1.42$.

\section{\texorpdfstring{$\SO(N_c)$ SQCD}{SO(Nc) SQCD}}\label{sec.SO}

Let us now discuss the case of the $\SO(N_c)$ gauge group.
The duality for this case is worked out in \cite{Aharony:2013kma} (see \cite{Benini:2011mf,Hwang:2011ht,Aharony:2011ci} for a closely related discussion for $\O(N_c)$ theories).

It should be kept in mind that the details of the duality depends on the 
global properties of the gauge group (e.g.\ $\SO(N_c), \O(N_c), \textrm{Spin}(N_c)$ or $\textrm{Pin}(N_c)$),
as well as the set of local operators we include to the theory.
For example, for the $\O(N_c)$ gauge group, we have two choices, which are denoted by $\O(N_c)_{\pm}$ \cite{Aharony:2013kma}.
In the following we first deal with the case of the
$\SO(N_c)$ gauge group, and then come back to the cases of $\O(N_c)_{+}, \O(N_c)_{-}, {\rm Spin}(N_c)$ and $\textrm{Pin}(N_c)$
gauge groups in section \ref{subsec.OSpinPin}.

\subsection{Dual Pairs}

\paragraph{Electric Theory}
The electric theory has
quarks $Q$ in the fundamental representation,
and as usual we have $W_{\rm electric}=0$.
The theory also has the monopole operator $Y$,
the baryon $B$, as well as a composite ``baryon-monopole operator'' $\beta$.

The theory has 
$\SU(N_f) \times \U(1)_A \times \U(1)_{\rm R-VU}$ 
continuous flavor symmetry,
under which the fields $Q, \tilde{Q}, Y$ transform as follows:
\begin{align}\label{SO_elec}
\begin{tabular}{c||c|cccccc}
 & $\SO(N_c)$ & $\SU(N_f)$ & $\U(1)_A$ & $\U(1)_{\rm R-UV}$  & $\mathbb{Z}_2^{\mathcal{C}}$ & $\mathbb{Z}_2^{\mathcal{M}}$ & $\mathbb{Z}_2^{\tilde{\mathcal{M}}}$ \\
\hline
\hline
$Q$ & $\bm{N_c}$ & $\bm{N_f}$ & $1$ & $0$   & & &  \\
\hline
$M$ & $\bm{1}$ & $\bm{N_f(N_f+1)/2}$ & $2$ & $0$ &$+1$ &$+1$&$+1$   \\
$Y$ & $\bm{1}$ & $\bm{1}$ & $-N_f$ & $N_f-N_c+2$ &$+1$ &$-1$ &$-1$ \\
$B$ & $\bm{1}$ & $(\bm{N_f})_A^{N_c}$ & $N_c$ & $0$  &$-1$ &$+1$&$-1$  \\
$\beta$ & $\bm{1}$ & $(\bm{N_f})_A^{N_c-2}$ & $-(N_f-N_c+2)$ & $N_f-N_c+2$ &$-1$&$-1$  &$+1$
\end{tabular}
\end{align}
Here $(\bm{N_f})_A^{N_c-2}$ and  $(\bm{N_f})_A^{N_c}$ represents totally antisymmetric representations.
We have listed there discrete symmetry $\mathbb{Z}_2^{\mathcal{C}}$, $\mathbb{Z}_2^{\mathcal{M}}$
and $\mathbb{Z}_2^{\tilde{\mathcal{M}}}$ (we here list charges only for gauge-invariant fields).
We can easily check that $\mathbb{Z}_2^{\tilde{\mathcal{M}}}$ is a combination of $\mathbb{Z}_2^{\mathcal{C}}$
and $\mathbb{Z}_2^{\mathcal{M}}$, and is not independent.
These discrete symmetries will play crucial roles when we change the gauge groups later in section \ref{subsec.OSpinPin}.

\paragraph{Magnetic Theory}

Let us consider the case $N_f>N_c+1$.
First, the magnetic theory has dual quarks $q$.
The meson $M$ as well as the monopole operator $Y$
of the electric theory, 
are now fundamental fields in the magnetic theory.
The magnetic theory also has the monopole operator
$\tilde{Y}$, the baryon $\tilde{B}$, and the dual baryon-monopole $\tilde{\beta}$.

Note that  the baryon $B$ and the baryon-monopole $\beta$
do not appear as fundamental fields of the magnetic theory (compare this with the case of $\SU(N_c)$ theory).
Rather they are identified with their magnetic counterparts $\tilde{B}, \tilde{\beta}$,
by an identification 
\begin{align}
(B, \beta) \longleftrightarrow (\tilde{\beta}, \tilde{B}) \ .
\label{Exchange}
\end{align}

The magnetic theory has a superpotential 
\begin{align}
\label{SO_W_1}
W_{\rm magnetic}=\frac{1}{2} M q q +  \frac{i^{N_f-N_c}}{4} \tilde{Y} Y \ . 
\end{align}

The theory has the same flavor symmetries as the electric theory,
under which the fields transform as follows:
\begin{align}\label{SO_mag}
\begin{tabular}{c||c|cccccc}
 & $\SO(N_f-N_c+2)$ & $\SU(N_f)$ & $\U(1)_A$ & $\U(1)_{R-UV}$ &$\mathbb{Z}_2^{\mathcal{C}}$ &$\mathbb{Z}_2^{\mathcal{M}}$  &$\mathbb{Z}_2^{\tilde{\mathcal{M}}}$ \\
\hline
\hline
$q$ & $\bm{N_f-N_c+2}$ & $\overline{\bm{N_f}}$ & $-1$ & $1$ & & &\\
$M$ & $\bm{1}$ & $\bm{\frac{1}{2} N_f(N_f+1)}$ & $2$ & $0$  & $+1$& $+1$& $+1$ \\
$Y$ & $\bm{1}$ & $\bm{1}$ & $-N_f$ & $N_f-N_c+2$  & $+1$& $-1$ & $-1$\\
\hline
$\tilde{Y}$ & $\bm{1}$ & $\bm{1}$ & $N_f$ & $N_c-N_f$  & $+1$& $-1$& $-1$\\
$\tilde{B}$ & $\bm{1}$ & $(\bm{N_f})_A^{N_c}$ & $-(N_f-N_c+2)$ & $N_f-N_c+2$ & $-1$& $-1$ & $+1$\\
$\tilde{\beta}$ & $\bm{1}$ & $(\bm{N_f})_A^{N_c-2}$ & $N_c$ & $0$  & $-1$& $+1$& $-1$
\end{tabular}
\end{align}
Note that the charge assignment for $\mathbb{Z}_2^{\mathcal{M}}$
and $\mathbb{Z}_2^{\tilde{\mathcal{M}}}$ symmetries 
here is consistent with the identification \eqref{Exchange}.

For the case $N_f=N_c-1$ the magnetic theory has no gauge group,
contains fields $Y$ and $M$, with superpotential given by (see \cite{Benini:2011mf,Hwang:2011ht,Aharony:2011ci}
for $\O(N_c)_{+}$ case)
\begin{align}\label{SO_W_2}
W = Y^2 \textrm{det}\, (M) + M q q \;.
\end{align}
The charge assignment is given by
\begin{align}
\begin{tabular}{c||ccc}
 &  $\SU(N_f)$ & $\U(1)_A$ & $\U(1)_{R-UV}$ \\
\hline
\hline
$q$ &  $\bm{\overline{N_f}}$ & $-1$ & $1$  \\
$M$ &  $\bm{\frac{1}{2} N_f(N_f+1)}$ & $2$ & $0$  \\
$Y$ &  $\bm{1}$ & $-N_f$ & $1$ 
\end{tabular}
\end{align}
and as before this case should be treated separately from the rest.

For lower values of $N_f$,
we have the quantum-corrected moduli space for $N_f= N_c-2$ \cite{Aharony:2011ci},
and the supersymmetry is broken for $N_f<N_c-2$.
We will hereafter concentrate on the case $N_f\ge N_c-1$.

\subsection{IR Analysis}

Let us parametrize the IR R-symmetry by
\begin{align}\label{IRsymm-SO}
R_{\rm IR}=R_{\rm UV}+ a J_A \ ,
\end{align}
where as before $R_{\rm UV}, R_{\rm IR}, J_A$ are generators of 
$\U(1)_{\rm R-UV}, \U(1)_{\rm R-IR}, \U(1)_A$, respectively.

\paragraph{Unitarity Bound}

Let us consider the electric theory.
The dimensions of the operators are given by
\begin{align}
\begin{split}
&M: \, \Delta_M=2 a  \  ,\\
&Y: \, \Delta_Y=(N_f-N_c+2) - N_f \, a \ ,  \\
&B: \, \Delta_B=N_c\, a  \ , \\
&\beta: \, \Delta_{\beta}=(N_f-N_c+2)-a(N_f-N_c+2) \ .
\end{split}
\end{align}
The unitary bound $\Delta\ge \frac{1}{2}$ gives
\begin{align}\label{SOunitarity}
\begin{split}
&M: \,  a\ge \frac{1}{4}   \ , \\
&Y: \, a\le \frac{N_f-N_c+\frac{3}{2}}{N_f}\approx 1-\frac{1}{x} \ ,  \\
&B: \, a\ge \frac{1}{2N_c} \ ,  \\
&\beta: \, a\le \frac{N_f-N_c+\frac{3}{2}}{N_f-N_c+2}\approx 1  \ .
\end{split}
\end{align}
Notice that in the Veneziano limit, the unitarity bound for the monopole operator $Y$ depends on the value of $x$,
whereas that for the baryon-monopole $\beta$ is independent of $x$.

\paragraph{Partition Function}
Let us write down the $S^3$ partition functions of electric  and magnetic theories.
The precise expression depends on whether $N_c$ is even or odd.

For $N_c$ even with $N_c=2 r$, the electric partition function is given by (see \eqref{Cartan} and \eqref{roots})
\es{ElectricPF-SOeven}{
\begin{split}
Z_{\rm electric}&=
{1 \over 2^{r} r!}\int \prod_{i=1}^{r} d\sigma_i   \, 
\prod_{1\le i<j\le r} [2\sinh[\pi(\sigma_i+\sigma_j)]2\sinh[\pi(\sigma_i-\sigma_j)]]^2 \\
&\quad \times
\overbrace{\prod_{i=1}^r \exp\left[
  N_f \, l(1-a \pm i \sigma_i ) 
\right]}^{Q} \ , 
\end{split}
}
while $N_c$ odd with $N_c=2 r+1$, we have
\es{ElectricPF-SOodd}{
\begin{split}
Z_{\rm electric}&=
{1 \over 2^{r} r!}\int \prod_{i=1}^{r} d\sigma_i   \, 
\prod_{1\le i<j\le r} [2\sinh[\pi(\sigma_i+\sigma_j)]2\sinh[\pi(\sigma_i-\sigma_j)]]^2 \\
&\quad \times
\prod_{i=1}^r
[2\sinh[\pi \sigma_i]]^2
\overbrace{\prod_{i=1}^r \exp\left[
  N_f \, l(1-a \pm  i \sigma_i ) 
\right] \ .
}^{Q}
\end{split}
}

The magnetic partition function is similar, and 
for $N_f-N_c+2$ even  ($N_f-N_c+2=:2\tilde{r}$)
\es{MagneticPF-SOeven}{
\begin{split}
Z_{\rm magnetic}&=
{1 \over 2^{\tilde{r}}\tilde{r}!}\exp\left[\overbrace{l\left(1-(N_f-N_c+2)+N_f a\right)}^{Y}+\overbrace{\frac{N_f(N_f+1)}{2} l(1-2a)}^{M}\right]\\
&\int \prod_{i=1}^{\tilde{r}} d\sigma_i  
\prod_{1\le i<j\le \tilde{r}} [2\sinh[\pi(\sigma_i+\sigma_j)]2\sinh[\pi(\sigma_i-\sigma_j)]]^2 \\
&\quad \times
\overbrace{
 \prod_{i=1}^r \exp\left[
  N_f \, l(a \pm i \sigma_i ) 
\right] 
}^{q} \ .
\end{split}
}
For $N_f-N_c+2=:2\tilde{r}+1$ with $N_f> N_c+1$ (i.e.\ $\tilde{r}>0$), we have
\es{MagneticPF-SOodd}{
\begin{split}
Z_{\rm magnetic}&=
{1 \over 2^{\tilde{r}}\tilde{r}!} \exp\left[\overbrace{l\left(1-(N_f-N_c+2)+N_f a\right)}^{Y}+\overbrace{\frac{N_f(N_f+1)}{2} l(1-2a)}^{M}\right] \\
&\int \prod_{i=1}^{\tilde{r}} d\sigma_i  
\prod_{1\le i<j\le \tilde{r}} [2\sinh[\pi(\sigma_i+\sigma_j)]2\sinh[\pi(\sigma_i-\sigma_j)]]^2 \\
&\quad \times
\prod_{i=1}^{\tilde{r}} [2\sinh(\pi\sigma_i)]^2 
\times 
\overbrace{
 \prod_{i=1}^{\tilde{r} } \exp\left[
  N_f \, l(a \pm i \sigma_i ) 
\right] }^{q}
\ .
\end{split}
}
For $N_f=N_c-1$ (i.e.\ $\tilde{r}=0$), we have
\es{MagneticPF-SOodd_2}{
\begin{split}
Z_{\rm magnetic}&=
\exp\left[\overbrace{l\left(N_f a\right)}^{Y}+\overbrace{\frac{N_f(N_f+1)}{2} l(1-2a)}^{M}
+\overbrace{N_f \, l(1-a)}^{q} \right] 
\ .
\end{split}
}

The convergence bounds of the partition functions
are given in the following form, which hold irrespective of whether $N_c, N_f-N_c+2$
are even or odd:
\es{Convergebd-SO}{
&\textrm{electric: } \quad a<  \frac{N_f-N_c+2}{N_f} \approx 1-\frac{1}{x}  \ ,  \\
&\textrm{magnetic: } \quad a> 
\frac{N_f-N_c}{N_f}
\approx 1-\frac{1}{x}  \ .
}
In this case there is a small overlapping region where
both electric and magnetic descriptions are valid. However the width of the overlapping region
shrinks to zero in the Veneziano limit.
It is therefore expected that we really should not expect both electric and magnetic descriptions to be valid,
except for only for limited values of $N_c$ and $N_f$.

\begin{figure}[h]
\begin{center}
  \includegraphics[scale=1.0]{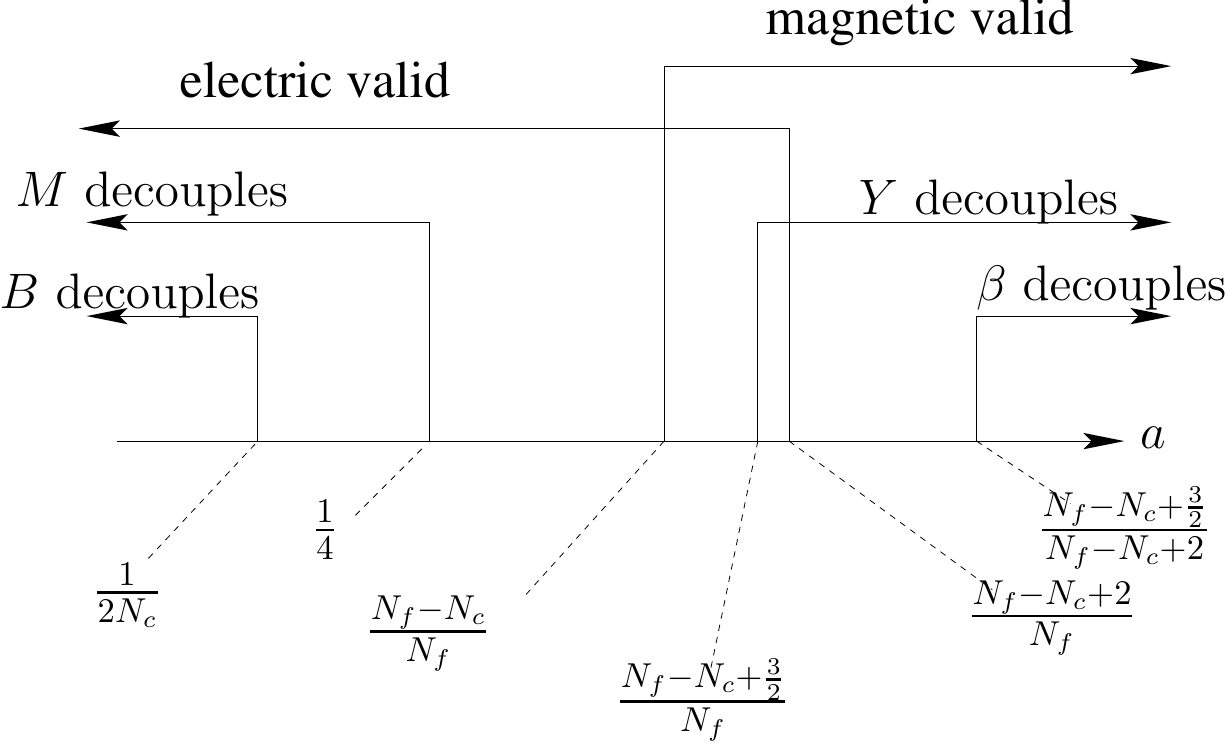}
  \caption{The unitarity bound and the convergence bound for 3d $\mathcal{N}=2$ $\SO(N_c)$ SQCD with $N_f$ flavors with $N_f> N_c-1$, plotted in terms of the mixing parameter $a$ (see \eqref{IRsymm-SO}). The correct IR value of $a$ should be determined from $F$-maximization.
 Depending on the value of $a$, operators decoupling might be either none, only $Y$, or both $Y$ and $\beta$. While the baryon-monopole $\beta$ could in principle decouple, this does not happen in the exampled we studied, both numerically and analytically.}
  \label{fig.SO}
  \end{center}
  \end{figure}

\paragraph{$F$-maximization}

We have done the $F$-maximization for several small values of $N_c$ and $N_f$.\footnote{Note that part of the entries have appeared in \cite{Hwang:2011ht}, which gives the critical values of $a$ consistent with ours, in the first $F$-maximization. In several cases, however, the monopole operators decouple and we need to do another $F$-maximization, to determine the correct $\U(1)$ R-symmetry.}

\begin{table}[h]
\begin{center}
\begin{tabular}{| c | cccccc |}
 \hline
    & $N_f = 1$ & $N_f = 2$ & $N_f = 3$ & $N_f = 4$ & $N_f = 5$  & $N_f = 6$\\
   \cline{1-7} $N_c = 2$  & \spc{$1/3$ \\ $(.8724)$ } & \spc{$.4085$ \\ $(2.6271)$ } &  \spc{$.4370$ \\ $(3.5311)$ }  & \spc{$.4519$ \\ $(4.3723)$} &  \spc{$.4611$ \\ $(5.1795)$} &  \spc{$.4674$ \\ $(5.9655)$} \\
  $N_c = 3$ &  - & \fcolorbox{red}{white}{\spc{$.2775$ $[Y]$ \\$(1.8721)$} } & \spc{$.3532$ \\ $(3.8607)$ } & \spc{$.3923$ \\$(5.4647)$ } &  \spc{$.4149$ \\ $(6.9098)$ } &  \spc{$.4297$ \\ $(8.2670)$ } \\
  $N_c = 4$ & - & - & \fcolorbox{red}{white}{\spc{$.2324$ $[Y]$ \\$(2.9315)$}} & \spc{$.3175$ \\ $(6.7624)$} & \spc{$.3602$ \\ $(9.1754)$} & \spc{$.3865$ \\ $(11.3076)$} \\
 $N_c = 5$ & - & - & - & \fcolorbox{red}{white}{\spc{$.2663$ $[Y]$\\ $(4.7671)$}} & \fcolorbox{blue}{white}{\spc{$.2909$ \\ $(8.7275)$}} & \spc{$.3358$ \\ $(11.8361)$}\\
  $N_c = 6$ & - & - & - & - & \fcolorbox{red}{white}{\spc{$.2635$ $[Y]$ \\$(6.7334)$}} & \fcolorbox{blue}{white}{\spc{$.2686$ \\ $(11.3107)$}} \\
 \hline
  \end{tabular}
\end{center}
\caption{The scaling dimension $\Delta_Q$ of the flavor multiplets (above) and the maximal value of $F$-function (below), at the conformal fixed points for a few small values of $N_f$ and $N_c$ in the 3d ${\cal N} = 2$ $\SO(N_c)$ (or ($\O_{+}(N_c)$) SQCD with $N_f$ flavors.
For the diagonal entries ($N_f=N_c-1$) we have used the partition function of the magnetic theory,
and for entries in red box we need to decouple the monopole operator $Y$.
All other entries are computed in the electric theory, except in blue boxes and in diagonal entries where we used the magnetic theory for better numerical computations.
}
\label{SOtable}
\end{table}

\begin{figure}[htbp]   
\leavevmode
\begin{center}
\scalebox{.78}{\includegraphics{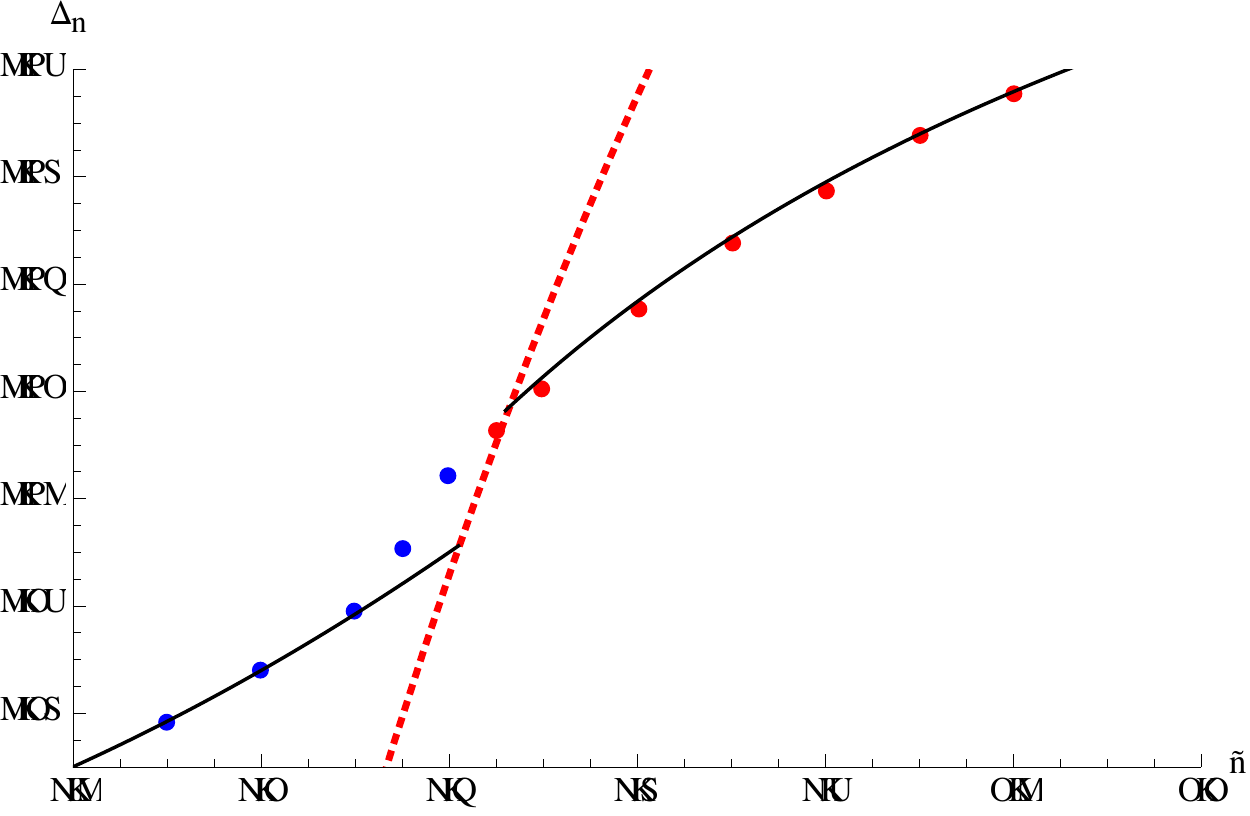}}
\end{center}
\caption{$\Delta_Q$ as a function of $x=N_f/N_c$ in the Veneziano limit. Points were computed by extrapolating small $N_c$ numerical results. Dotted line is the unitarity bound~\eqref{SOunitarity}. We find that $\Delta_Q(x)$ hits the unitarity bound at the critical value $x_c \approx 1.45$. The black curves at large and small values of $x$ are the analytical approximation~\eqref{DeltaSOlimit} and~\eqref{DeltaSO_mag}, respectively. In the region right to the red curve, we use electric theory, while in the left region we use magnetic theory with monopole $Y$ decoupled.}
\label{fig: CrackSO}
\end{figure}

As commented before, one interesting feature of $\SO(N_c)$ theories is the existence of the 
baryon-monopole operator $\beta$. This means that the baryon-monopole $\beta$,
in addition to the baryon $B$, could decouple in the IR.
In the examples we studied in Table \ref{SOtable}, however,
we find that $\beta$ never decouples (this is also the case in the Veneziano limit, to be discussed below, see Figure \ref{fig: CrackSO}).

We also compute $\Delta_Q(N_c, N_f)$ for both $N_c$ odd and even,
in the large $N_f$ limit.
The $S^3$ partition functions take slightly different forms for $N_c$ odd and even,
however it is natural to think that 
the value of $\Delta_Q(N_c, N_f)$ should coincide between the two cases 
in this limit. Analytic calculation by order in ${1 \over N_f}$ shows that odd/even cases give the same answer,
which reads
\es{DeltaSO}{
\Delta_Q(N_c, N_f) =  {1 \over 2} - {2N_c - 2 \over \pi^2 N_f} + {(24N_c^2 - 48N_c + 24) - \pi^2 ({10 \over 3}N_c^2 -8N_c +{14 \over 3}) \over \pi^4 N_f^2}+ \mathcal{O}\left({1 \over N_f^3}\right) \,.
}
In the Veneziano limit this expansion reduces to
\es{DeltaSOlimit}{
\Delta_Q(x) = {1 \over 2} - {2 \over \pi^2 x} + {8(36 - 5\pi^2) \over 12\pi^4 x^2}  + \mathcal{O}\left({1 \over x^3}\right) \,. 
}
In the case of the magnetic theory, we can expand $\Delta(N_c, N_f)$ in order of ${1 \over N_c}$ which gives us
\es{}{
\Delta_Q(N_c, N_f) = {1 \over 4} + {1 \over 4\pi}{N_f - N_c + 2 \over N_c} + {(N_f - N_c +2)^2 \over 8\pi^2(\pi-2)} + O\left({1 \over N_c^3}\right)  \ , 
}
which reduces in the Veneziano limit into 
\es{DeltaSO_mag}{
\Delta_Q(x) = {1 \over 4} + {1 \over 4\pi} (x-1) + {(26 - 7\pi)\pi - 8 \over 8\pi^2 (\pi-2)} (x-1)^2 + O((x-1)^3) \ .
}

\subsection{\texorpdfstring{$\O(N_c)_{\pm}$, ${\rm Spin}(N_c)$ and ${\rm Pin}(N_c)$ Gauge Groups}{O(Nc), Spin(Nc) and Pin(Nc) Gauge Groups}}\label{subsec.OSpinPin}

\paragraph{Dual Pairs}

Let us now come to the other gauge groups, $\O(N_c)_{\pm}$, ${\rm Spin}(N_c)$ and ${\rm Pin}(N_c)$.
Note that all these gauge groups have the same Lie algebra as that of $\SO(N_c)$.
These dualities can be obtained by gauging discrete $\mathbb{Z}_2$ symmetries of the $\SO(N_c)$ electric and magnetic theories. 

When we gauge a single $\mathbb{Z}_2$ symmetry, there are three choices:
 $\mathbb{Z}_{\mathcal{C}}$, $\mathbb{Z}_{\mathcal{M}}$
and $\mathbb{Z}_{\tilde{\mathcal{M}}}$, leading to $\O(N_c)_{+}$, $\O(N_c)_{-}$ and ${\rm Spin}(N_c)$ 
theories, respectively, for the electric theory.

We can apply the same gauging to the magnetic theory.
In fact, all the terms which appear in the magnetic superpotentials (see \eqref{SO_W_1} and \eqref{SO_W_2})
have charge $+1$ under any of the three $\mathbb{Z}_2$ symmetries,
and hence the gauging is consistent  with the superpotential.
There is one big difference from the electric case, however: the role of the $\mathbb{Z}_{\mathcal{M}}$
and $\mathbb{Z}_{\tilde{\mathcal{M}}}$ should be exchanged, as follows from the identification \eqref{Exchange}.

When we gauge two $\mathbb{Z}_2$ symmetries, there is only one choice,
since we are gauging all the discrete symmetries, and we obtain the ${\rm Pin}(N_c)_{-}$ theory.

\begin{figure}[htbp]
\begin{center}
\begin{tikzcd}[row sep=large, column sep=huge]
%
 & \fcolorbox{blue}{white}{$\O(N_c)_{-}$} \arrow{dr}{\mathbb{Z}_2^{\mathcal{C}}}& \\
\fcolorbox{red}{white}{$\SO(N_c)_{+}$}  \arrow{ur}{\mathbb{Z}_2^{\tilde{\mathcal{M}}}}  \arrow{r}{\mathbb{Z}_2^{\mathcal{C}}} \arrow{dr}{\mathbb{Z}_2^{\mathcal{M}}} & \fcolorbox{red}{white}{$\O(N_c)_{+}$} \arrow{r}{\mathbb{Z}_2^{\mathcal{M}} \textrm{ or } \mathbb{Z}_2^{\tilde{\mathcal{M}}}} & \fcolorbox{blue}{white}{${\rm Pin}(N_c)$} \\
 & \fcolorbox{blue}{white}{${\rm Spin}(N_c)$} \arrow{ur}{\mathbb{Z}_2^{\mathcal{C}}}& 
\end{tikzcd}
\end{center}
\caption{By gauging either $\mathbb{Z}_{\mathcal{C}}$, $\mathbb{Z}_{\mathcal{M}}$
and $\mathbb{Z}_{\tilde{\mathcal{M}}}$ symmetries of the $\SO(N_c)$ theory we obtain
$\O(N_c)_{+}$, $\O(N_c)_{-}$ and ${\rm Spin}(N_c)$ gauge groups.
Further gauging the remaining $\mathbb{Z}_2$, we obtain the ${\rm Pin}(N_c)$ theory.
This figure represents the gauging of the electric theory, and for the magnetic theory
we should exchange the role of $\mathbb{Z}_{\mathcal{M}}$
and $\mathbb{Z}_{\tilde{\mathcal{M}}}$.
As we will explain later, the gauge groups boxed in red and those boxed in blue 
have different monopole operators, differing by a power of $2$ (see \eqref{Yspin}).
}
\label{fig.OSpinPin}
\end{figure}
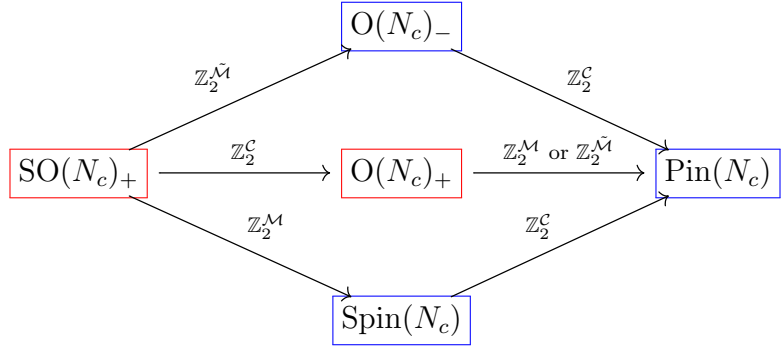

These gauging patterns are summarised in Figure \ref{fig.OSpinPin}.

This immediately implies that 
the correct Aharony-like duality works as \cite{Aharony:2013kma}
\begin{align}
\begin{split}
&\O_{+}(N_c) \longleftrightarrow \O_{+}(N_f-N_c+2) \ ,\\
&\O_{-}(N_c) \longleftrightarrow {\rm Spin}(N_f-N_c+2) \ ,\\ 
&{\rm Spin}(N_c) \longleftrightarrow \O_{-}(N_f-N_c+2) \ , \\ 
&{\rm Pin}(N_c) \longleftrightarrow {\rm Pin}(N_f-N_c+2) \ . 
\end{split}
\end{align}
This should be compared with the $\SO$ duality
\begin{align}
\SO(N_c) \longleftrightarrow \SO(N_f-N_c+2) \ . 
\end{align}

When we gauge discrete $\mathbb{Z}_2$ symmetries,
we project out the fields which has charge $-1$.

For example, when gauging the $\mathbb{Z}_2^{\mathcal{C}}$ symmetry (charge conjugation symmetry)
we obtain the dualities for $\O_{+}$ gauge groups.
In this case, the baryon $B$ and the baryon-monopole $\beta$ are projected out,
while their combinations, such as $B^2$ and $B\beta$, remain in the theory.

Similarly, when we gauge either $\mathbb{Z}_{\mathcal{M}}$ or $\mathbb{Z}_{\tilde{\mathcal{M}}}$
symmetry, the monopole operator in itself is projected out, and we instead have its square remaining:
\begin{align}\label{Yspin}
Y_{\rm Spin}=Y_{\O_{-}} =Y_{\rm Pin}:=Y^2 \ .
\end{align}

\paragraph{IR Analysis}

We now come to a natural question: does the gauging of the discrete symmetries discussed above have any impact on the
IR behavior of the theory?

It turns out most of the preceding analysis for $\SO$ gauge groups does not require any modification.
This is because we are primarily interested in $F$-maximization, 
which requires only the $S^3$ partition function with no operators inserted,
and hence is insensitive to gauging of discrete symmetries.

There is one big change, however. While we have the same set of operators, gauging makes
some of the gauge-invariant operators gauge-non-invariant.
Since the unitarity bound applies only to gauge-invariant operators,
the discrete symmetry gauging will in general change the unitarity bounds.

In the analysis for the $\SO$ gauge groups we did not find any examples where the baryon $B$,
the meson $M$, or the baryon-monopole $\beta$ decouple. We can therefore concentrate on the 
monopole operator $Y$. As we discussed above, the change for ${\rm Spin}$, $\O_{-}$ and ${\rm Pin}$ gauge groups is that 
the gauge-invariant monopole operator is not $Y$, but rather $Y_{\rm spin}=Y^2$ \eqref{Yspin},
whose scaling dimension is twice that of $Y$.

This immediately means that the unitarity bound for $Y$ in \eqref{SOunitarity}
is replaced by
\begin{align}
\begin{split}
&Y_{\rm spin}: \, a\le \frac{N_f-N_c+\frac{7}{4}}{N_f}\approx 1-\frac{1}{x}  \ ,  \\
\end{split}
\end{align}
As expected, the difference from the discrete symmetry gauging goes away in the Veneziano limit.

We can redo the IR decoupling analysis, to obtain the new table as in Table \ref{Spintable}.
Clearly the only difference can happen when the monopole operator $Y$ decouples in the $\SO$ theory.
In the table this happens when $N_f=N_c-1=2, 3$, when the monopole operator $Y$ no longer decouples.

\begin{table}
\begin{center}
\begin{tabular}{| c | cccccc |}
 \hline
    & $N_f = 1$ & $N_f = 2$ & $N_f = 3$ & $N_f = 4$ & $N_f = 5$  & $N_f = 6$\\
   \cline{1-7} $N_c = 2$  & \spc{$1/3$ \\ $(0.8724)$ } & \spc{$.4085$ \\ $(2.6271)$ } &  \spc{$.4370$ \\ $(3.5311)$ }  & \spc{$.4519$ \\ $(4.3723)$} &  \spc{$.4611$ \\ $(5.1795)$} &  \spc{$.4674$ \\ $(5.9655)$} \\
  $N_c = 3$ &  - & \spc{$.2696$ \\$(1.8664)$} & \spc{$.3532$ \\ $(3.8607)$ } & \spc{$.3923$ \\$(5.4647)$ } &  \spc{$.4149$ \\ $(6.9098)$ } &  \spc{$.4297$ \\ $(8.2670)$ } \\
  $N_c = 4$ & - & - &\spc{$.2324$ \\$(2.9316)$} & \spc{$.3175$ \\ $(6.7624)$} & \spc{$.3602$ \\ $(9.1754)$} & \spc{$.3865$ \\ $(11.3076)$} \\
 $N_c = 5$ & - & - & - & \fcolorbox{red}{white}{\spc{$.2663$ $[Y]$\\ $(4.7671)$}} & \fcolorbox{blue}{white}{\spc{$.2909$ \\ $(8.7275)$}} & \spc{$.3358$ \\ $(11.8361)$}\\
  $N_c = 6$ & - & - & - & - & \fcolorbox{red}{white}{\spc{$0.2635$ $[Y]$ \\$(6.7334)$}} & \fcolorbox{blue}{white}{\spc{$.2686$ \\ $(11.3107)$}} \\
 \hline
  \end{tabular}
\end{center}
\caption{The scaling dimension $\Delta_Q$ of the flavor multiplets (above) and the maximal value of $F$-function (below), at the conformal fixed points for a few small values of $N_f$ and $N_c$ in the 3d ${\cal N} = 2$ ${\rm Spin}(N_c)$ ($\O_{-}(N_c)$ or ${\rm Pin}(N_c)$) SQCD with $N_f$ flavors.
For the diagonal entries ($N_f=N_c-1$) we have used the partition function of the magnetic theory,
and for entries in red box we need to decouple the monopole operator $Y$.
All other entries are computed in the electric theory, except in blue boxes and in the diagonal entries where we used the magnetic theory for better numerical computations. Compare this with Table \ref{SOtable}.
}
\label{Spintable}
\end{table}

\section{Digression on Group Theory}\label{sec.group}

While the analysis in sections \ref{sec.SU}, \ref{sec.USp} and \ref{sec.SO} are treated separately,
some of the structures are can understood in a more unified manner from the representation theory of the Lie algebras,
as one might expect.

To give one example for this phenomena, let us point out that 
the scaling dimension of the quarks and anti-quarks $\Delta_{Q}=\Delta_{\tilde{Q}}$, in the large $N_f$ expansion
up to the order of ${1 \over N_f^2}$, can be uniformly represented as
\es{Delta}{
\Delta_Q(N_c, N_f)= {1 \over 2} - {4 C_F \over \pi^2}{1 \over N_f} + {96C_F^2 - \pi^2(8C_F^2 + {8 \over 3}C_F C_A) \over \pi^4}{1 \over N_f^2} + O\left( {1 \over N_f^3} \right) \,,
}
where $C_F$ and $C_A$ are quadratic Casimirs in fundamental and adjoint representation, respectively. 
Concretely, we have
\begin{align}
\begin{split}  
\U(N_c):&\quad C_F=\frac{N_c}{2} \ ,  \quad C_A= N_c - {1 \over N_c} \ , \\
\SU(N_c):&\quad  C_F=\frac{N_c^2-1}{2 N_c} \ , \quad C_A= N_c \ , \\
\USp(2N_c):&\quad C_F=\frac{2 N_c+1}{4} \ , \quad C_A= N_c+1 \ , \\
\SO(N_c): &\quad  C_F=\frac{N_c-1}{2} \ , \quad  C_A= N_c-2 \ ,  
\end{split}
\label{C_FA}
\end{align}
with which we can verify that \eqref{Delta} reproduce the formulas \eqref{DeltaSU}, \eqref{DeltaUSp}, \eqref{DeltaSO} and \eqref{DeltaU}.
We expect the formula \eqref{Delta} to apply to SQCD with other gauge groups, e.g.\ exceptional gauge groups.
That the quadratic Casimirs appear in the expansion coefficients can be explained by the fact that the coefficients 
compute the certain Feynman diagrams ({\it cf.} \cite[Appendix B]{Safdi:2012re}).
Note that in the leading Veneziano limit we always have
$C_F\sim \frac{N_c}{2}, C_A\sim N_c$ and the differences of the gauge groups are washed away.
This is basically the reason that the plots of $\Delta_Q$ (in Figures \ref{fig:CrackSU}, \ref{fig: CrackUSp} and 
\ref{fig: CrackSO}),
as well as the critical value $x_c$, are similar among different choices of gauge groups with the same ranks.

\section{Gauging and Quiver Gauge Theories}\label{sec.quiver}

The difference between $\U(N_c)$ theory and the $\SU(N_c)$ theory is an example 
where the gauging of a flavor symmetry dramatically modifies the IR dynamics.
We have also seen in section \ref{subsec.OSpinPin} that gauging of discrete symmetries
also changes the IR decoupling.
These can be thought of as particular examples of more general phenomena where
the gauging of a flavor symmetry modifies the IR dynamics of the theory.

As yet another example of this type, we
study gauging of the $\SU(N_f)$ flavor symmetry
of the SQCD, to obtain a quiver gauge theory with a product gauge group.
We discuss the effect of the gauging to the IR R-symmetry, and to the decoupling of monopole operators.
Such quiver gauge theories naturally arise in string theory (see e.g\ \cite{He:2002cr,Yamazaki:2008bt} and references therein),
and (as we will discuss later in the next section)
for example in the 
compactifications of M5-branes.

\subsection{Electric Gauging}

Let us start with the $\U(N_c)$ SQCD with $N_f$ flavors, discussed in section \ref{app.UN}.

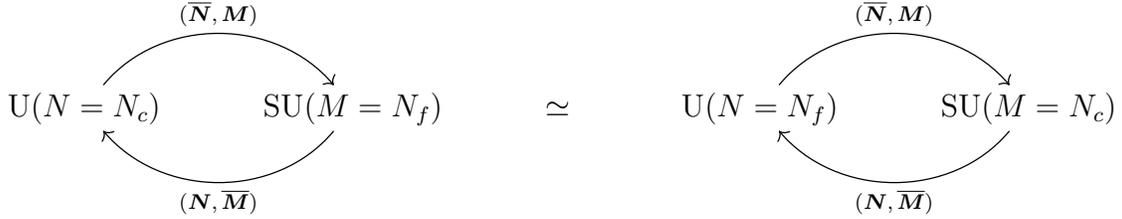
\begin{figure}[htbp]
\begin{center}
\begin{tikzcd}
\U(N=N_c)
  \arrow[bend left=50]{r}[name=U]{(\overline{\bm{N}}, \bm{M})}
 &
\SU(M=N_f)
 \arrow[bend left=50]{l}[name=D]{(\bm{N}, \overline{\bm{M}})} 
 & 
 \simeq
 &
\U(N=N_f)
  \arrow[bend left=50]{r}[name=U]{(\overline{\bm{N}}, \bm{M})}
 &
\SU(M=N_c)
 \arrow[bend left=50]{l}[name=D]{(\bm{N}, \overline{\bm{M}})} 
\end{tikzcd}
\end{center}
\caption{A Quiver diagram for the electric quiver gauge theory considered in this section.
The quiver gauge theory can be obtained by gauging the $\SU(M)$ flavor symmetry of 
$\U(N)$ SQCD with $M$ flavors, or by gauging the $\U(N)$ flavor symmetry of $\SU(M)$ SQCD with $N$ flavors.
Note that the gauge group can be thought of $\SU(N)\times \U(M), \U(N)\times \SU(M)$ or 
$(\U(N)\times \U(M))/\U(1)$, all describing the same theory.
The theory therefore has symmetry exchanging $N$ and $M$.
The arrows represent bifundamental $\mathcal{N}=2$ chiral multiplets.
}
\label{fig.quiverNM}
\end{figure}

As shown in \eqref{tab.U_electric}, this theory has $\SU(N_f)_L\times \SU(N_f)_R$
symmetry. Let us choose to gauge the diagonal $\SU(N_f)$ of these two $\SU(N_f)$ symmetries,
which we denote by $\SU(N_f)_V$. The resulting theory then has 
$\U(N_c)\times \SU(N_f)$ gauge symmetry,
and $\SU(N_f)_A\times \U(1)_A\times \U(1)_J\times \U(1)_{\rm R-UV}$ flavor symmetry,
where $\SU(N _f)_A$ is the axial (anti-diagonal) combination of $\SU(N_f)_L\times \SU(N_f)_R$.
As a result of this gauging, we obtain a quiver gauge theory, whose quiver diagram is shown in 
Figure \ref{fig.quiverNM}.

One remark is that the resulting theory has a symmetry exchanging 
$N_c$ and $N_f$. In fact, the gauge group can equivalently be taken as
$(\U(N_c)\times \U(N_f))$, with matters in the bifundamental representations
$Q: (\bm{N_c}, \overline{\bm{N_f}})$ and $\tilde{Q}: (\overline{\bm{N_c}}, \bm{N_f})$.
This is a quiver gauge theory, and since we only have bifundamental matters
the overall $\U(1)$ of the $(\U(N_c)\times \U(N_f))$ gauge group trivially decouples,
leading to the $(\U(N_c)\times \U(N_f)) / U(1)\simeq \U(N_c)\times \SU(N_f)$ gauge group as before.

In the rest of this section we use the notation 
$N:=N_c, M:=N_f$, to make this symmetry more manifest.
In fact, now that the symmetry is manifest we can regard the same quiver gauge theory as obtained from the 
gauging $\SU(N)_V$ flavor symmetry of the $\U(M)$ SQCD with $N$ flavors, with $N$ and $M$ reversed from above (Figure \ref{fig.quiverNM}).

Now, the question we ask in this subsection is whether or not this gauging of the $\SU(M)$ symmetry
has any effect in the discussion of the IR scaling dimension and the decoupling of monopole operators.

The best way to see this is to write down the $S^3$ partition function as the 
parameter $a$ corresponding to the mixing of the $\U(1)_A$ symmetry, as in Appendix \ref{app.UN}:
\es{Quiver_SUNM_1}{
\begin{split}
Z_{\U(N) \times \SU(M)}(a)&=
{1 \over N!} {1 \over M!}\int \prod_{i=1}^{N} d\sigma_i  \prod_{i=1}^{M} d\rho_j  , \delta\left(\sum_{j=1}^M \rho_j\right)   \,  \\
&\quad \times \overbrace{\prod_{1\le i<j\le N} \sinh^2[\pi(\sigma_i-\sigma_j)] \prod_{1\le i<j\le M} \sinh^2[\pi(\rho_i-\rho_j)]}^{\rm measure}\\
&\quad \times
\overbrace{\prod_{i=1}^{N} \prod_{i=1}^{M}\exp\left[
   l(1-a+ i \sigma_i -i\rho_j ) + l(1-a- i \sigma_i +i \rho_j) 
\right]}^\textrm{bifundamental $Q, \tilde{Q}$} \ .
\end{split}
}
Note that the integral is kept invariant under the simultaneous shift of $\sigma_i$ and $\rho_j$,
and this represents the overall decoupled $\U(1)$ commented before. The same partition function can also be written as
\es{Quiver_eg_1}{
\begin{split}
Z_{(\U(N)\times \U(M))/\U(1)}(a)&=
{1 \over N!} {1 \over M!}\int \prod_{i=1}^{N} d\sigma_i  \prod_{i=1}^{M} d\rho_j 
\, \delta\left(\sum_{i=1}^N \sigma_i+ \sum_{j=1}^M \rho_j\right) \\
  \, 
&\quad   \times
\overbrace{\prod_{1\le i<j\le N} \sinh^2[\pi(\sigma_i-\sigma_j)] \prod_{1\le i<j\le M} \sinh^2[\pi(\rho_i-\rho_j)]}^{\rm measure}\\
&\quad \times
\overbrace{\prod_{i=1}^{N} \prod_{i=1}^{M}\exp\left[
   l(1-a+ i \sigma_i -i\rho_j ) + l(1-a- i \sigma_i +i \rho_j)  
\right]}^\textrm{bifundamental $Q, \tilde{Q}$} \ .
\end{split}
}

In either way, it is clear that gauging dramatically change
the partition function as a function of the parameter $a$,
and consequently the IR R-charges/conformal dimensions of the theory.
In fact, this is to be expected since we have a manifest symmetry between $N$ and $M$ after gauging;
by contrast $N_c=N, N_f=M$ theory and $N_c=M, N_f=N$ clearly have different IR dynamics, 
as we have seen in the rest of this paper.

This symmetry between $N$ and $M$ is actually a source for trouble,
when we consider the convergence bound for the electric $S^3$ 
partition function. The convergence bound before gauging was worked out in \eqref{Convergebd1},
and since now we have symmetry and $N$ and $M$, we should impose the same constraint 
with $N$ and $M$ ($N_c$ and $N_f$) exchanges.
This gives
\begin{align}
a< \textrm{min} \left(
\frac{M-N+1}{M} \ , \quad \frac{N-M+1}{N}
\right) \le \textrm{min} \left(
\frac{1}{M} \ , \,\, \frac{1}{N}
\right)
\ ,
\end{align}
and in particular $a$ will be negative unless $N=M$. 

Note that convergence constraint is ameliorated by including flavor matters to gauge groups $\SU(N)$ and $\SU(M)$.
Suppose that we include $k$ flavors ($l$ flavors) to gauge groups $\SU(N)$ and $\SU(M)$.
The convergence constraint then reads 
\begin{align}
a< \textrm{min} \left(
\frac{M-N+k+1}{M} \ , \quad \frac{N-M+l+1}{N}
\right) \ ,
\end{align}
and in particular the constraint in practice goes away for sufficiently large $k$ and $l$.
Such flavors are natural from string theory constructions, however
we will set $k=l=0$ in the discussion below, to simplify analysis.

\subsection{Magnetic Gauging}

To avoid this convergence issue,
one might be tempted to switch to the magnetic description.
Namely, instead of gauging the flavor symmetry of the electric theory,
we can choose to gauge the flavor symmetry of the magnetic theory.

For the case $M>N\ge 1$, we can first go to the magnetic theory of $\SU(N)$ SQCD with $M$ flavors,
go to the magnetic dual, and then gauge the $\U(M)\simeq \SU(M)\times U(1)_B$ flavor symmetry of the theory.
Note that these magnetic descriptions break the symmetry between $N$ and $M$, as is clear from the fact that the 
magnetic descriptions requires us to take $M\ge N$.
The resulting partition function is given by (compare this with \eqref{MagneticPF1}):
\es{Quiver_mag_2}{
Z_{\rm magnetic}^{\SU(N)\times \U(M)}& =
\frac{1}{ (M-N)!}{1 \over M!}\int \prod_{i=1}^{M-N} d\sigma_i  \prod_{i=1}^{M} d\rho_j \, d\rho \,\,
\delta\left(\sum_{j=1}^M \rho_j \right) 
\, 
\\
& \times
\overbrace{\prod_{1\le i<j \le M-N} \sinh^2[\pi(\sigma_i-\sigma_j)] \prod_{1\le i<j\le M} \sinh^2[\pi(\rho_i-\rho_j)]}^{\rm measure} \, e^{2\pi b \rho}\\
&\times\overbrace{\prod_{i,j=1}^M \exp\left[ \, l(1-2a +i\rho_i- i \rho_j)\right]}^{M} 
\exp\left[ \overbrace{ l\left(1+2M \, a-2(M-N+1) \right) }^{Y}\right]  \\
&\times \overbrace{ \prod_{i=1}^{M-N}  \prod_{j=1}^{M} 
\exp\left[  
 l\left(a\pm  i \sigma_i  \mp i\rho_j \right)
\right] }^{q,  \, \tilde{q}} \\
&
\times \overbrace{\exp\left[  
     l\left(1+(M-N)-M \, a \pm iN\, \rho \mp i  \sum_{1\le j\le M-N} \sigma_j \right)
    \right] 
    }^{ b, \, \tilde{b}}
 \ .   
}
Equivalently, we can start with the $\U(N)$ SQCD with $M$ flavors,
and then gauge the $\SU(M)$ flavor symmetry (compare \eqref{MagneticPF1-U}):
\es{Quiver_mag_3}{
\begin{split}
Z_{\rm magnetic}^{\U(N)\times \SU(M)}& =
{1 \over (M-N)!}{1 \over M!}
\overbrace{\exp\left[  l\left(1-(M-N+1)\pm b+M \, a \right) \right]}^{V_{\pm}}\\
&\times
\int \prod_{i=1}^{M-N} d\sigma_i \prod_{j=1}^{M} d\rho_j  \, \delta\left(
\sum_{j=1}^{M} \rho_j
\right) \\
&\times
\overbrace{\prod_{1\le i<j\le M-N} \sinh^2[\pi(\sigma_i-\sigma_j)]  
 \prod_{1\le i<j\le M} \sinh^2[\pi(\rho_i-\rho_j)]}^{\rm measure} \, 
\\
&\times \overbrace{ \prod_{i,j=1}^{M} \exp\left[ \, l(1-2a+i \rho_i - i \rho_j)  \right] }^{M} \\
&\times \overbrace{\prod_{i=1}^{M-N} 
\prod_{j=1}^{M} 
\exp\left[  
 l\left(a\pm i(  \sigma_i  - \rho_j) \right)
\right] }^{q, \, \tilde{q}} \ .
\end{split}
}
The formal equivalence of the two expressions \eqref{Quiver_mag_2} and \eqref{Quiver_mag_3}
(up to a constant phase factor)
can be checked directly by 
using the identities \eqref{XYZdual} and \eqref{lodd}. (In fact, this is essentially the argument
used for deriving $\SU$ dualities from $\U$ dualities, as reviewed in Appendix \ref{app.UtoSU}).

Unfortunately, the convergence bound for 
\eqref{Quiver_mag_2} and \eqref{Quiver_mag_3} is satisfied for 
\begin{align}
\begin{split}
a> \frac{M-N-1}{M} \quad \textrm{and } \quad a<0 \ ,
\end{split}
\end{align}
where the first (second) inequality comes from the convergence of the $\sigma$ ($\rho$) integrals.
In other words the partition function \eqref{Quiver_mag_2} and \eqref{Quiver_mag_3} 
cannot be used for any practical $F$-maximization.

The situation is better for 
the case $M=N$. We can then gauge the $\SU(M=N)$ flavor symmetry of the magnetic $\U(N)$ theory, 
leading to the partition function
\begin{align}\label{MN1}
\begin{split}
Z^{N=M}_{\rm magnetic}
&={1 \over N!} \overbrace{\exp\left[  l(N a\pm b )\right]}^{V_{\pm}} 
\prod_{i=1}^{N} \int d\rho_j 
\, \delta\left(\sum_{j=1}^N \rho_j\right) \overbrace{\prod_{i,j=1}^N \exp\left[l(1-2a+ i\rho_i -i\rho_j)\right]}^M \ .
\end{split}
\end{align}
We can instead choose to gauge the $\U(N)\simeq \SU(N)_V\times U(1)_B$ flavor symmetry
of the magnetic $\SU(M)$ theory, leading to the expression
\begin{align}\label{MN2}
\begin{split}
Z^{N=M}_{\rm magnetic}
&={1 \over N!}\overbrace{  \exp\left[l(1-2+2N a )\right] }^{Y}
\int d\rho \, e^{2\pi b\rho} \, \overbrace{\exp\left[l(1-Na \pm i\rho) \right]}^{B, \, \tilde{B}} \\
& \times
\int  \prod_{j=1}^{N}  d\rho_j \, \delta\left(\sum_{j=1}^N \rho_j \right)
\, \overbrace{\prod_{i,j=1}^N \exp\left[ l(1-2a+ i\rho_i -i\rho_j)\right]}^M\ .
\end{split}
\end{align}
The equivalence of \eqref{MN1} and \eqref{MN2} can again be checked by 
using the identities \eqref{XYZdual} and \eqref{lodd}.
The convergence bound for these expressions is an inequality
\begin{align}
a< \frac{1}{2} \ .
\end{align}

It is natural to expect that the magnetic quiver theories discussed here are dual to the electric quiver gauge theories discussed before.
There is one caveat, however. We have implicitly assumed that the order of two operations, namely 
gauging of the flavor symmetry and going to the dual magnetic description, commute with each other. 
Since the duality at hand is an IR duality, 
in general the gauging of the flavor symmetry could change the behavior under the RG flow,
and hence spoil the IR duality. 

We however expect that this does not happen when the gauge coupling for the 
newly-gauged flavor symmetry is much smaller than other gauge couplings. If in the UV the gauge coupling for $\SU(N)$ is much larger than that for the $\SU(M)$ gauge group, one imagines that the strong coupling effect of $\SU(N)$ kicks in first, and then the dynamics of $\SU(M)$ does not matter until at much lower scales.
We can then discuss the dynamics of $\SU(N)$ and $\SU(M)$ gauge groups separately.
Since the $S^3$ partition function is independent of gauge couplings,
the equality of the $S^3$ partition functions should hold for any value of the gauge coupling,
giving further evidence for the duality after gauging.

\paragraph{Numerical Results}

The numerical analysis of the quiver case is computationally more challenging than the SQCD case,
and as we have seen the convergence bound tends to be severe.
Therefore let us here consider the simplest case of $N=M$ magnetic theory. We can then do  $F$-maximization
for the partition function \eqref{MN1} (compare with Table \ref{SUtable}). 
The numerical results for the values $N=M=2,3,4,5$ are summarized
in Table \ref{Quivertable}. 

There are two remarks on this result. First, the value of $a$ at the maximum is different from 
that before gauging, as expected. Another non-trivial result here is that here in none of these cases
exhibit operator decoupling. This is partly because the meson $M$ of the magnetic theory,
after gauging, is now an adjoint field with respect to the newly-introduced gauge symmetry,
and hence is not gauge invariant. Therefore there is no need to consider the 
unitarity bound of the meson itself.

\begin{table}
\begin{center}
\begin{tabular}{|  cccc |}
 \hline
    $N= 2$ & $N= 3$ & $N = 4$ & $N= 5$  
    \\
    \hline
  \spc{$.2172$ \\ $(3.5322)$ } & \spc{$.1962$  \\ $(7.5009)$ } &  \spc{$.1804$ \\ $(12.1339)$ }  
  & \spc{$.1567$ \\ $(15.2490)$ }  
  \\
  \hline
\end{tabular}
\end{center}
\caption{The critical value of the parameter $a$ (above) and the critical value of the $F$-function (below), for the $\SU(N)\times \U(N)\sim (\U(N)\times \U(N))/ \U(1)$ theory,
as computed from the magnetic partition function \eqref{MN1}.
Notice that the critical value of $a$ is different from that in Table \ref{SUtable}
before gauging the $\U(N)$ flavor symmetry. In all these cases there is no indication that 
any operator decouple in the IR.
}
\label{Quivertable} 
\end{table}

\subsection{General Quivers}

\begin{figure}[htbp]
\begin{center}
  \includegraphics[scale=0.8]{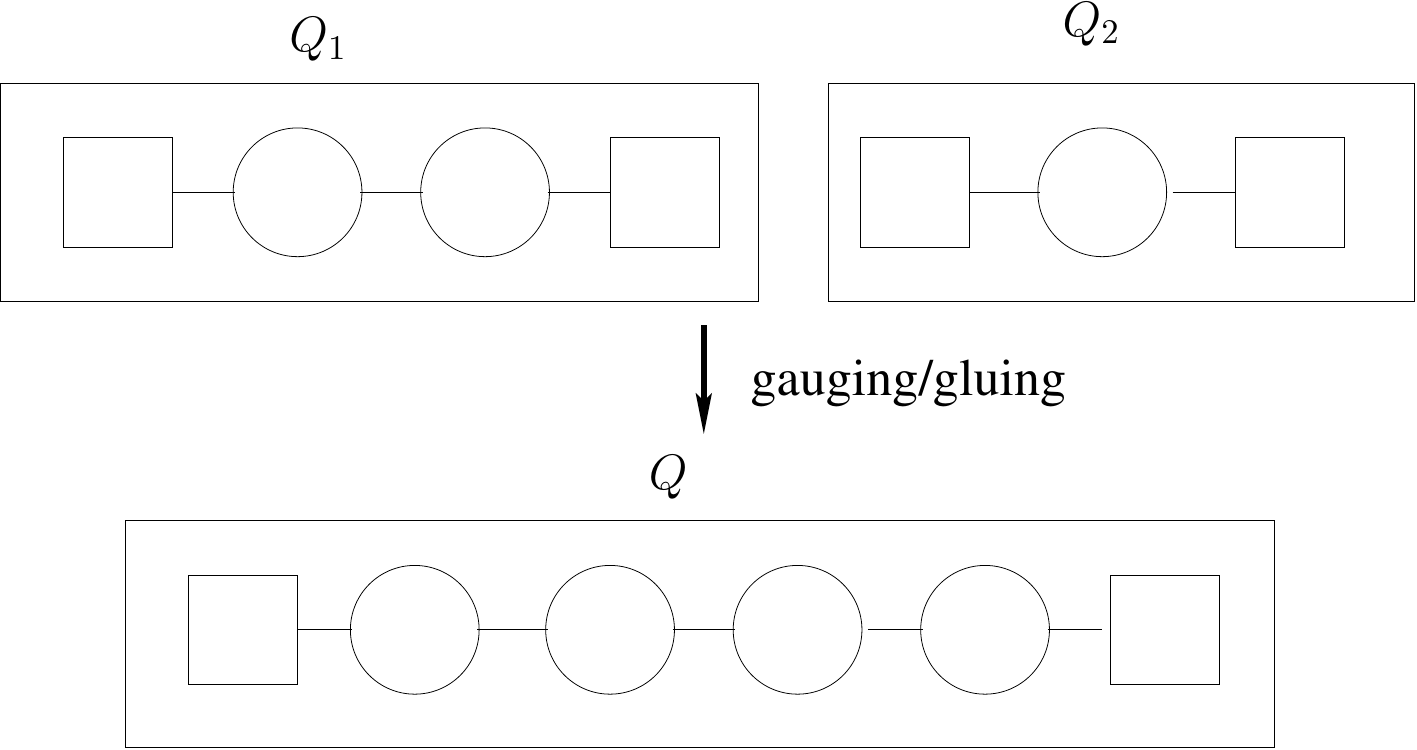}
  \caption{We can generate a larger quiver $Q$ by gluing together two quivers $Q_1$ and $Q_2$.
  In gauge theory language the circle (the square) represents the gauge (global) symmetry, each of which can be
  for example $\U(N_c)$ or $\SO(N_c)$ with different values of $N_c$ for different nodes.
   Gluing in this context means to take two flavor symmetries (represented by two squares in the middle, which we assume to contain the same flavor symmetry group)  and gauge the diagonal subgroup of the product. The partition function behaves nicely under this gluing,
   however not the $F$-maximization nor the IR behavior.}
  \label{fig.quiver}
  \end{center}
  \end{figure}

Having discussed quiver gauge theories with two nodes, we can discuss more general 3d $\mathcal{N}=2$ 
quiver gauge theories, whose matter content is determined by a quiver diagram, i.e.\ an oriented graph\footnote{We also need to specify the superpotentials, however this choice does not really modify the qualitative features of the conclusions below.}.
We can then gauge the appropriate flavor symmetries, whose effect is to 
concatenate two quiver diagrams and to generate a more complicated quiver diagram (Figure \ref{fig.quiver}).
For example, if we glue two quivers $Q_1, Q_2$ at a node
to obtain a new quiver $Q$, then partition function for the 
larger quiver $Q$ can be schematically written in the form
\begin{align}
Z_{Q}[\sigma_1, \sigma_3](a_1,a_2)= 
\int d\sigma_2 \,  Z_{\rm vector}[\sigma_2] \, Z_{Q_1}[\sigma_1, \sigma_2](a_1)\, Z_{Q_2}[\sigma_2, \sigma_3](a_2)  \ ,
\end{align}
where $Z_{\rm vector}[\sigma_2]$ is the contribution from the vector multiplet which is gauged under the gluing,
and $a_{1,2}$ denote the parameters repressing the flavor symmetries of the theories $Q_{1,2}$.
  
As before, the extremum of $Z_{Q}[\sigma_1, \sigma_3](a_1, a_2)$
as a function of $a_1$ ($a_2$) is in general 
different from that of the $Z_{Q_1}[\sigma_1, \sigma_2](a_1)$ ($Z_{Q_2}[\sigma_2, \sigma_3](a_2)$).
This means that to tell whether the monopole operator for the gauge group in the quiver $Q_1$ 
decouples or not, we need to know in advance the detailed data for the quiver $Q_2$,
however large the quiver $Q_2$ may be.\footnote{In the spirit of \cite{Yamazaki:2013xva} one might be tempted to say that there is a ``long-range entanglement in the theory space''.
It would be interesting to explore this point further, and connect the discussion here to the 
entanglement in the dual statistical mechanical model discussed in \cite{Yamazaki:2012cp,Terashima:2012cx}.} This is in sharp contrast with the 
case of 3d $\mathcal{N}=4$ supersymmetry, where the 
IR decoupling of the monopole operators can be checked locally at the quiver diagram, by verifying the 
inequality $N_f\ge 2 N_c$ ({\it cf.} \cite{Cottrell:2015jra,Cottrell:2016nsu} for recent discussion in gravity dual).

\section{Implications for M5-brane Compactifications}\label{sec.M5}

The comments from the previous section
has interesting implications to the M5-brane compactifications,
which we now turn to.

\subsection{\texorpdfstring{Boundary Conditions of 4d $\mathcal{N}=4$ SYM}{Boundary Conditions of 4d N=4 SYM}}

Let us first start with the results of \cite{Gaiotto:2008ak},
which classifies $1/2$--BPS boundary conditions of 4d $\mathcal{N}=4$ Super Yang-Mills Theory (SYM).
Some of the boundary conditions (of the Neumann type, realized by D3-branes ending on NS5-branes in type IIB string theory)
contain a non-trivial boundary field theory $\mathfrak{B}$, which is given as a 3d $\mathcal{N}=4$ linear-chain quiver gauge theory.
When $\mathfrak{B}$ contains monopole operators decoupling in the IR, we regard the corresponding 
$\mathfrak{B}$ as containing redundant degrees of freedom, in the sense that $\mathfrak{B}$ contains fields which decouples completely from the bulk 4d $\mathcal{N}=4$ theory.
Since we are interested in the classification of the minimal set of boundary conditions in the IR, 
this means we can disregard such $\mathfrak{B}$ from the classification of boundary conditions.
This, together with the criterion $N_f\ge 2 N_c$ mentioned above, lead to the conclusion that the 
choices of $\mathfrak{B}$ are exhausted by the so-called $T_{\rho}[\SU(N)]$ theories, 
with $\rho$ being the partition of $N$.\footnote{
If we consider the mixture of Dirichlet and Neumann boundary conditions we obtain a slightly more general class of theories $T_{\rho}^{\sigma}[\SU(N)]$, 
labeled by a pair of partitions $\rho, \sigma$.
}

We can now consider the $1/4$--BPS boundary conditions \cite{Hashimoto:2014vpa,Hashimoto:2014nwa}, 
whose boundary field theory $\mathfrak{B}$ would then
be 3d $\mathcal{N}=2$ theories. As we have seen above the criterion for the decoupling/un-decoupling
is now more complicated than the inequality $N_f\ge 2 N_c$. In particular, in the Veneziano limit\footnote{
This is natural in the context of the holographic dual.} we have learned from \cite{Safdi:2012re}
that the decoupling happens at the critical value $x_c\approx 1.45<2$. This suggests that 
the minimal set of (Neumann type) boundary conditions should no longer be labeled by partitions.\footnote{
In 3d $\mathcal{N}=2$ theories, for each vertex of the quiver diagram we have the choice of 
whether or not to include $\mathcal{N}=2$ adjoint chiral multiplet. This means that the natural generalization of the 
$1/2$--BPS analysis is that the boundary conditions are labeled by a decorated partition.
However, our point here is that this is likely a redundant characterization of the IR boundary condition.
}
It would be interesting to see if/how this conclusion could fit together with the 
analysis of the generalized Nahm equations in \cite{Hashimoto:2014vpa,Hashimoto:2014nwa},
or their 4d $\mathcal{N}=1$ counterparts \cite{Bonelli:2013pva,Xie:2013gma}.

\subsection{\texorpdfstring{Co-dimension $2$ Defects of 6d $(2,0)$ Theory}{Co-dimension 2 Defects of 6d (2,0) Theory}}

Since the 4d $\mathcal{N}=4$ $\SU(N)$ SYM is an $S^1$ compactification of 5d $\mathcal{N}=2$ $\SU(N)$ SYM (super-Yang-Mills) and also
a $T^2$ compactification of the 6d $(2,0)$ theory of $A_{N-1}$ type,
we can try to lift the conclusions of the previous subsection into the statements on the 5d $\mathcal{N}=2$ SYM and the 6d $(2,0)$ theory.

For the $1/2$-BPS case, the same consideration leads to the conclusion that (see \cite[section 2.1]{Gang:2015wya} for review)
the $1/4$--BPS co-dimension 2 defects are labeled by a partition $\rho$ of $N$, and the 
effect of the defect in the 5d language is to couple the 5d $\mathcal{N}=2$ SYM to the $T_{\rho}[\SU(N)]$ theory mentioned above. 
This fact has been utilized recently in the context of the compactifications of the 6d theories on
2-manifolds \cite{Yonekura:2013mya,Bullimore:2014upa,Bullimore:2014awa}
and 3-manifolds \cite{Gang:2015bwa,Gang:2015wya}.

Now we can repeat the same argument for the $1/4$--BPS boundary conditions, and again obtain $1/4$--BPS boundary conditions for 
5d $\mathcal{N}=2$ SYM and the 6d $(2,0)$ theory. Our conclusion is then that these defects should {\it not} be labeled by partitions, 
since otherwise we would be over-counting. This should have some interesting counterparts as data specifying $1/4$--BPS defects in 
4d $\mathcal{N}=2$ theories arising from the 2-manifold compactifications \cite{Gaiotto:2009we,Wyllard:2009hg}, or 
3d $\mathcal{N}=2$ theories arising from the 3-manifold compactifications \cite{Terashima:2011qi,Dimofte:2011ju,Lee:2013ida,Cordova:2013cea}.

\section*{Acknowledgements}
We would like to thank I.\ Klebanov for encouragement, advice and for careful readings of this manuscript. We also thank I.\ Yaakov and B.\ Willett for discussion.
J. L.\ received support from the Samsung Scholarship.
M. Y.\ was supported by WPI program (MEXT, Japan), by JSPS Program for Advancing Strategic International Networks to Accelerate the Circulation of Talented Researchers, by JSPS KAKENHI Grant No.\ 15K17634, and by Institute for Advanced Study. 


\appendix

\section{Monopole Operators}\label{app.monopole}

Given a root $\alpha$ of the gauge group, we can write down the corresponding monopole operator
\begin{align}
Y_{\alpha} \simeq \exp\left( \frac{\alpha(\sigma) }{g_3^2} +i \alpha(a) \right) \ ,
\end{align}
where  $(\sigma_1, \ldots, \sigma_r)$ are the Cartan part of the scalar in the adjoint vector multiplet,
and $(a_1, \ldots, a_r)$ are the dual photon of the Cartan part of the gauge group.
The latter is periodic with period $2\pi g_3^2$, making $Y_{\alpha}$ well-defined.

Only the $Y_{\alpha}$'s for positive simple roots $\alpha$ are independent, and hence classically
we have $r$ independent monopole operators. This parametrizes the classical Coulomb branch.
However many of these Coulomb branches are lifted by quantum corrections
(instanton-generated superpotential).

For example, for $\SU(N_c)$, classically we have $N_c-1$ monopole operators
\begin{align}
Y_j \simeq \exp\left( \frac{\sigma_j-\sigma_{j+1} }{g_3^2} +i (a_j-a_{j+1}) \right) \ ,
\quad j=1, \ldots, N_c-1 \ ,
\end{align}
however the only remaining operator in the end is 
\begin{align}
Y=\prod_{j=1}^{N_c-1} Y_j\simeq \exp\left( \frac{\sigma_1-\sigma_{N_c} }{g_3^2} +i (a_1-a_{N_c}) \right) \ .
\end{align}
This is the monopole operator discussed in section \ref{sec.SU}.

\section{\texorpdfstring{$S^3$ Partition Functions}{S3 Partition Functions}}\label{app.S3}

The $S^3$ partition function \cite{Kapustin:2009kz,Jafferis:2010un,Hama:2010av} is given by
(in the absence of Chern-Simons terms, FI parameters and real mass parameters)
\begin{align}
Z=\frac{1}{|W|}\int_{\rm Cartan} \!\!d \sigma \,\, 
\prod_{\alpha: \textrm{root}} \left[\sinh\pi \alpha(\sigma)\right]^2
\prod_{\Phi: \, \textrm{chiral multiplet}}
\prod_{\rho: \, \textrm{weight of } R_{\Phi}}
 \exp\left[ l\left(1-\Delta_{\Phi}+i \rho(\sigma) \right) \right] \ ,
 \label{S3}
\end{align}
where $|W|$ is the order of the Weyl group,
$R_\Phi$ ($\Delta_{\Phi}$) is the 
representation under the gauge group (R-charge) of the chiral multiplet $\Phi$
and the function $l(z)$ is defined by
\begin{align}
l(z):=-z \log(1-e^{2\pi i z})
+\frac{i}{2} \left(\pi z^2+\frac{1}{\pi}
\textrm{Li}_2(e^{2\pi i z})\right)  
-\frac{i\pi}{12} \ .
\end{align} 
This function $l(z)$ has poles at integers on the real axis, except at the origin.
We also have the relation
\begin{align}\label{lodd}
l(x)+l(-x)=0 \ .
\end{align}
For convergence of the partition function we use the following asymptotics
in the limit $\sigma_1\to \infty$:
\es{Expansion1}{
&l(a \mp i \sigma_1) = \pm {i \pi \over 2} \sigma_1^2 - \pi a \sigma_1 + \mathcal{O}(\sigma_1^0) \ ,\\
&\prod_{i<j}^{N_c} \sinh^2[\pi(\sigma_i - \sigma_j)] = e^{2\pi(N_c - 1) \sigma_1 + \mathcal{O}(\sigma_1^0)}\,.
}
For $\USp/\SO$ gauge groups the Cartan subalgebra is parametrized as
\es{Cartan}{
& \USp(2r): \,  \{   \sigma_1, \ldots \sigma_r, -\sigma_1, \ldots, -\sigma_r  \}  \ , \\
& \SO(2r ): \, \{  \sigma_1, \ldots \sigma_r, -\sigma_1, \ldots, -\sigma_r  \} \ , \\
& \SO(2r+1): \, \{  \sigma_1, \ldots \sigma_r, -\sigma_1, \ldots, -\sigma_r ,0 \} \ ,
}
and the roots are given by
\es{roots}{
& \USp(2r): \,  \{ \pm \sigma_i \pm \sigma_j \}_{1\le i<j\le r} \cup \{ \pm 2 \sigma_i \}_{1\le i\le r} \ ,\\
& \SO(2r ): \, \{\pm \sigma_i \pm \sigma_j\}_{1\le i<j\le r} \ , \\
& \SO(2r+1): \, \{\pm \sigma_i \pm \sigma_j\}_{1\le i<j\le r} \cup \{ \pm \sigma_i\}_{1\le i\le r} \ .
}

\section{\texorpdfstring{$\U(N_c)$ SQCD}{U(Nc) SQCD}}\label{app.UN}

In this Appendix we briefly summarize the case of the 
$\U(N_c)$ gauge group \cite{Safdi:2012re}.
It is instructive to compare the discussion in this Appendix with that of the
$\SU(N_c)$ theory in the main text.
Some of the ingredients discussed in this Appendix will be used in the discussion of quiver gauge theories in section \ref{sec.quiver}.

\subsection{Dual Pairs}

\paragraph{Electric Theory}

The electric theory is similar to the $\SU(N_c)$  SQCD.
The major difference is that we have 
two remaining monopole operators $V_{\pm}$.

The theory has a $\U(N_c)$ gauge symmetry, as well as
 $\SU(N_f)_L \times \SU(N_f)_R \times \U(1)_A \times \U(1)_J\times \U(1)_{\rm R-UV}$ flavor symmetries,
under which the fields $Q, \tilde{Q}, Y$ transform as follows:

\begin{align}
\centering
\begin{tabular}{c||c|ccccc}
 & $\U(N_c)$ & $\SU(N_f)_L$ & $\SU(N_f)_R$  & $\U(1)_A$  & $\U(1)_J$ & $\U(1)_{\rm R-UV}$ \\
\hline
\hline
$Q$ & $\bm{N_c}$ & $\bm{N_f}$ & $\bm{1}$  & 1  & 0 & 0 \\
$\tilde{Q}$ & $\overline{\bm{N_c}}$ & $\bm{1}$ & $\overline{\bm{N_f}}$ & 1 & 0  & 0 \\
\hline
$V_{\pm } $ & $\bm{1}$ & $\bm{1}$ & $\bm{1}$ & $-N_f$ & $\pm 1$ & $N_f-N_c+1$ \\
\end{tabular}
\label{tab.U_electric}
\end{align}
Note that compared with the $\SU(N_c)$ case we have the topological $\U(1)_J$ symmetry in this case,
whereas the $\U(1)_B$ symmetry, being part of the gauge symmetry, is absent.

\paragraph{Magnetic Theory}

Let us first assume $N_f>N_c$.
The magnetic theory has gauge group 
$\U(\tilde{N}_c)$ (remember the definition $\tilde{N}_c:=N_f-N_c$), and has
dual quark $q$, anti-quarks $\tilde{q}$,
and the meson $M=Q\tilde{Q}$ and $V_{\pm }$.
The magnetic theory also has two monopole operators
$\tilde{V}_{\pm}$ for the dual photons of magnetic gauge groups.
The superpotential is given by
\es{}{
W_{\rm magnetic}=\tilde{q} M q+V_{+} \tilde{V}_{-} + V_{-} \tilde{V}_{+} \ .
}

The theory again has the same flavor symmetry as the electric theory,
under which the fields transform as follows:
\begin{align}
\centering
\begin{tabular}{c||c|cccccc}
 & $\U(\tilde{N}_c)$ & $\SU(N_f)_L$ & $\SU(N_f)_R$  & $\U(1)_A$ & $\U(1)_J$&  $\U(1)_{\rm R-UV}$ \\
\hline
\hline
$q$ & $\bm{\tilde{N}_c}$ & $\overline{\bm{N_f}}$ & $\bm{1}$ & $-1$ &$0$ & $1$ \\
$\tilde{q}$ & $\overline{\bm{\tilde{N}_c}}$  & $\bm{1}$ & $\bm{N_f}$  & $-1$ &$0$ & 1 \\
$M$ & $\bm{1}$ & $\bm{N_f}$ & $\overline{\bm{N_f}}$  & $2$ &$0$  & $0$ \\
$V_{\pm}$ & $\bm{1}$ & $\bm{1}$ & $\bm{1}$  & $-N_f$  &$\pm 1$ & $\tilde{N}_c+1$ \\
\hline
$\tilde{V}_{\pm}$ & $\bm{1}$ & $\bm{1}$ & $\bm{1}$ & $N_f$  &$\pm 1$ & $-\tilde{N}_c+1$\\
\end{tabular}
\label{tab.U_magnetic}
\end{align}

For $N_f=N_c$, the magnetic theory do not have a gauge group,
and is described by the chiral superfields $V_{\pm}, M$, with the superpotential
\begin{align}
W=V_{+} V_{-} \textrm{det}(M) \ .
\label{VVM}
\end{align}
The charge assignment in this case is 
\begin{align}
\centering
\begin{tabular}{c||cccccc}
 & $\SU(N_f)_L$ & $\SU(N_f)_R$  & $\U(1)_A$ & $\U(1)_J$&  $\U(1)_{\rm R-UV}$ \\
\hline
\hline
$M$ &$\bm{N_f}$ & $\overline{\bm{N_f}}$  & $2$ &$0$  & $0$ \\
$V_{\pm}$ & $\bm{1}$ & $\bm{1}$  & $-N_f$  &$\pm 1$ & $\tilde{N}_c+1$ \\
\end{tabular}
\end{align}

\subsection{IR Analysis}

As in other cases discussed in the main text,
we need to consider the IR-mixing of the 
$\U(1)$ R-symmetry with the $\U(1)_A$ symmetry
\es{IRsymm1-U}{
R_{\rm IR}=R_{\rm UV}+ a J_A \ .
}
Note that we do not need to consider the mixing with the 
topological $\U(1)_J$ symmetry, since otherwise the parity is broken.

\paragraph{Unitarity Bound}

The dimensions of $V_{\pm}$ and $M$ are given by
\es{Dim1-U_new}{
\Delta_{V_{\pm}}=(N_f-N_c+1)-N_f a  \ ,\qquad
\Delta_M = 2a \ ,
}
which leads to the unitarity bound is given by
\es{Dim1-U}{
V_{\pm} : \, a \le \frac{N_f-N_c+\frac{1}{2}}{N_f}  \ ,\qquad
M: \, a\ge \frac{1}{4} \ ,
}
which in the Veneziano limit simplifies to 
\es{Venezianobd1-U}{
\frac{1}{4}\le a \le 1-\frac{1}{x} \ .
}
Note this requires $x\ge \frac{4}{3}$, and we will find the crack
before this value.

\paragraph{Partition Function}

The partition function of the electric theory is given by
\es{ElectricPF1-U}{
\begin{split}
Z_{\rm electric}&=
{1 \over N_c!}\int \prod_{i=1}^{N_c} d\sigma_i   \, 
\overbrace{\prod_{1\le i<j\le N_c} \sinh^2[\pi(\sigma_i-\sigma_j)] }^{\rm measure}\\
&\quad \times
\overbrace{\prod_{i=1}^{N_c} \exp\left[
  N_f \, l(1-a+ i \sigma_i ) + N_f\,  l(1-a- i \sigma_i ) 
\right]}^{Q, \, \tilde{Q}} \ ,
\end{split}
}
and that of the magnetic theory (for $N_f>N_c$) by
\es{MagneticPF1-U}{
\begin{split}
Z_{\rm magnetic}& =
{1 \over \tilde{N}_c!}\exp\left[ \overbrace{N_f^2 \, l(1-2a) }^{M}+\overbrace{2  l\left(1-(\tilde{N}_c+1)+N_f \, a \right)}^{V_{\pm}} \right]  \\
&\times
\int \prod_{i=1}^{\tilde{N}_c} d\sigma_i 
\overbrace{\prod_{1\le i<j\le \tilde{N}_c} \sinh^2[\pi(\sigma_i-\sigma_j)] }^{\rm measure} \, 
\overbrace{\prod_{i=1}^{\tilde{N}_c} 
\exp\left[  
    N_f \,  l\left(a+ i \sigma_i  \right)
  +    N_f \,  l\left(a- i \sigma_i \right)
\right] }^{q, \, \tilde{q}} \ .
\end{split}
}
The convergence of the expression for the partition function above gives
\es{Convergebd1-U}{
&\textrm{electric: } \quad a<  \frac{N_f-N_c+1}{N_f}  \approx 1-\frac{1}{x} \ ,  \\
&\textrm{magnetic: } \quad a> \frac{N_f-N_c-1}{N_f} \approx 1-\frac{1}{x}  \ .
}
As explained in the main text for the $\SU(N_c)$ case, we can either derive this from the positivity of the dimension of the monopole operators, or from the positivity of the effective FI parameter introduced in \eqref{zetaR}:
\es{Convergebd1-U-zeta}{
&\textrm{electric: } \quad \zeta_{\rm eff}=\overbrace{-(N_c-1)}^{\text{measure}}+\overbrace{N_f(1-a)}^{Q, \, \tilde{Q}}
=(N_f-N_c+1)-N_f\, a  \ ,  \\
&\textrm{magnetic: } \quad \zeta_{\rm eff}=\overbrace{-(\tilde{N}_c-1)}^{\text{measure}}+\overbrace{N_fa }^{q, \, \tilde{q}}  =-(N_f-N_c-1)+N_f\, a \  .
}

\begin{figure}[htbp]
\begin{center}
  \includegraphics[scale=1.0]{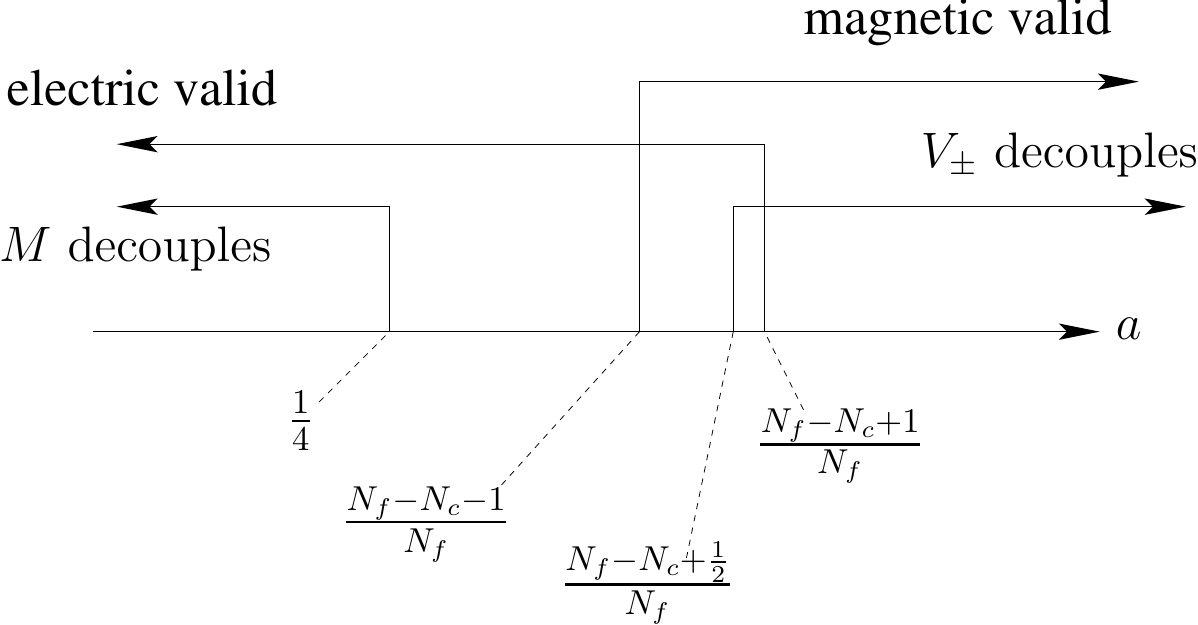}
  \caption{The unitarity bound and the convergence bound for the 3d $\mathcal{N}=2$ $\U(N_c)$ SQCD with $N_f$ flavors with $N_f> N_c$, plotted in terms of the mixing parameter $a$ (see \eqref{IRsymm1-U}). The correct IR value of $a$ should be determined from $F$-maximization.}
  \label{fig.U}
  \end{center}
  \end{figure}

The large $N_f$ and small $x-1$ expansions of the scaling dimensions of the matter quarks are given by
\begin{align}
\Delta_Q(N_c, N_f)= {1 \over 2} - {2 N_c \over \pi^2}{1 \over N_f} + {(24  - \pi^2 \frac{10}{3} )N_c^2  \over \pi^4}{1 \over N_f^2} + O\left( {1 \over N_f^3} \right) \,,
\label{DeltaU}
\end{align}
and 
\begin{align}
\Delta_Q(x)= \frac{1}{4}+ \frac{1}{4}(x-1)+ \frac{(26-7\pi)\pi-8}{8(\pi-2)\pi^2}(x-1)^2 + \mathcal{O}((x-1)^3) \ .
\end{align}

\section{\texorpdfstring{$\SU(N_c)$ Dualities from $\U(N_c)$ Dualities}{\SU(Nc) Dualities from \U(Nc) Dualities}}\label{app.UtoSU}

In this Appendix we derive the $\SU(N_c)$ dualities from the $\U(N_c)$ dualities.
The basic argument is not really new, and basically the same as in \cite{Park:2013wta}, except that here we work out the derivation
at the level of the $S^3$ partition function (as opposed to the 3d index in \cite{Park:2013wta}).
Similar manipulations appear in the discussion of quiver gauge theories in section \ref{sec.quiver}.

Let us begin with the partition functions of $\U(N_c)$ theories, 
with all the real mass/FI parameters to our partition functions turned on in
\eqref{ElectricPF1-U}, \eqref{MagneticPF1-U} (this means that $a$ is now complexified).
When we denote the real mass parameters for the $\U(1)_J, \SU(N_f)_L$ and $\SU(N_f)_R$ symmetries by
$\zeta, \mu_a, \tilde{\mu}_a$ ($a=1, \ldots, N_f$),  we have the $S^3$ partition functions
\es{ElectricPF1-U-all}{
\begin{split}
Z_{\rm electric}&=
{1 \over N_c!}\int \prod_{i=1}^{N_c} d\sigma_i   \, 
\,\, e^{-2\pi i \zeta \sum_{i=1}^{N_c} \sigma_i}
\prod_{1\le i<j\le N_c} \sinh^2[\pi(\sigma_i-\sigma_j)] \\
&\quad \times
\prod_{i=1}^{N_c} \prod_{a=1}^{N_f} \exp\left[
   l(1-a+ i \sigma_i +i \mu_a ) +   l(1-a- i \sigma_i +i \tilde{\mu}_a) 
\right] \ ,
\end{split}
}
and \es{MagneticPF1-U-all}{
\begin{split}
Z_{\rm magnetic}& =
{1 \over \tilde{N}_c!}\exp\left[ N_f^2 \, l(1-2a) +  l\left(1-(\tilde{N}_c+1)+N_f \, a\pm i \zeta \right) \right]  \\
&\times
\int \prod_{i=1}^{\tilde{N}_c} d\sigma_i 
\,\, e^{-2\pi i \zeta \sum_{i=1}^{\tilde{N}_c} \sigma_i}
\prod_{1\le i<j\le \tilde{N}_c} \sinh^2[\pi(\sigma_i-\sigma_j)]  
\\
&
\times 
\prod_{i=1}^{\tilde{N}_c} \prod_{a=1}^{N_f}
\exp\left[  
   l\left(a+ i \sigma_i  +i\mu_a\right)
  +    l\left(a- i \sigma_i +i \tilde{\mu}_a \right)
\right] \,.
\end{split}
}

Now to obtain the $\SU(N_c)$ duality all we need to do is apply the $S$-transformation
(as defined in \cite{Witten:2003ya}) to the $\U(1)_J$ global symmetry. In other words we add an off-diagonal Chern-Simons term
\begin{align}
\mathcal{L}=\frac{1}{4\pi} A_{\rm new} \wedge dA_{\U(1)_J} \ ,
\label{offCS}
\end{align}
and gauge the gauge field $A_{\U(1)_J}$ for $\U(1)_J$.
As we will see momentarily, the new gauge field $A_{\rm new}$
will be identified with that of the $\U(1)_B$ symmetry of the magnetic theory:
at the level of the $S^3$ partition function this amounts to the
Fourier transform with respect to $\zeta$.

For the electric theory, we have
\es{ElectricPF1-U-all-new}{
\begin{split}
Z'_{\rm electric}&=
{1 \over N_c!}\int d\zeta\,\, e^{2\pi  i N_c \hat{b}  \zeta} \int \prod_{i=1}^{N_c} d\sigma_i   \, 
\,\, e^{-2\pi i \zeta \sum_{i=1}^{N_c} \sigma_i}
\prod_{1\le i<j\le N_c} \sinh^2[\pi(\sigma_i-\sigma_j)] \\
&\quad \times
\prod_{i=1}^{N_c} \prod_{a=1}^{N_f} \exp\left[
   l(1-a+ i \sigma_i +i \mu_a ) +   l(1-a- i \sigma_i +i \tilde{\mu}_a) 
\right]\\
&=
{1 \over N_c!} \int \prod_{i=1}^{N_c} d\sigma_i   \, 
\,\, \delta\left(\sum_{i=1}^{N_c} \sigma_i-N_c \hat{b} \right) 
\prod_{1\le i<j\le N_c} \sinh^2[\pi(\sigma_i-\sigma_j)] \\
&\quad \times
\prod_{i=1}^{N_c} \prod_{a=1}^{N_f} \exp\left[
   l(1-a+ i \sigma_i +i \mu_a ) +   l(1-a- i \sigma_i +i \tilde{\mu}_a) 
\right]\\
&=
{1 \over N_c!}\int \prod_{i=1}^{N_c} d\sigma_i   \, 
\,\, \delta\left(\sum_{i=1}^{N_c} \sigma_i \right) 
\prod_{1\le i<j\le N_c} \sinh^2[\pi(\sigma_i-\sigma_j)] \\
&\quad \times
\prod_{i=1}^{N_c} \prod_{a=1}^{N_f} \exp\left[
   l\left(1-a+ i \sigma_i+i \hat{b} +i \mu_a \right) +   l\left(1-a- i \sigma_i -i \hat{b} +i \tilde{\mu}_a\right) 
\right] \ ,
\end{split}
}
where in the last line we shifted $\sigma_i \to \sigma_i+\sigma$.
We can check that this gives the charge assignment of the electric $\SU(N_c)$ theory,
and in particular that this answer gives the 
\eqref{ElectricPF1} when we take $\mu_a=\tilde{\mu}_a=0$
and when we identify $b=i \hat{b}$.\footnote{The factor of $i$ here is explained from the fact that the 
$S^3$ partition function depends on a complex combination, the real part $b$ being the 
anomalous dimension due to the mixing with a global symmetry
and the imaginary part being the real mass $\hat{b}$ \cite{Jafferis:2010un,Festuccia:2011ws}.}
Note also that in the Fourier transform we have included a factor of $N_c$; this was chosen such that the parameter $\hat{b}$
after the Fourier transform can be directly identified with the real mass parameter for the $\U(1)_B$ symmetry.

We can add the same off-diagonal Chern-Simons term \eqref{offCS}
to the magnetic theory, whose partition function is 
 \es{MagneticPF1-U-all-new}{
\begin{split}
Z'_{\rm magnetic}& =
{1 \over \tilde{N}_c!} \int d\zeta\,\, e^{2\pi  i N_c \hat{b} \zeta} 
\exp\left[ N_f^2 \, l(1-2a) +  l\left(1-(\tilde{N}_c+1)+N_f \, a\pm i \zeta \right) \right]  \\
&\times
\int \prod_{i=1}^{\tilde{N}_c} d\sigma_i 
\,\, e^{-2\pi i \zeta \sum_{i=1}^{\tilde{N}_c} \sigma_i}
\prod_{1\le i<j\le \tilde{N}_c} \sinh^2[\pi(\sigma_i-\sigma_j)]  
\\
&
\times 
\prod_{i=1}^{\tilde{N}_c} \prod_{a=1}^{N_f}
\exp\left[  
   l\left(a+ i \sigma_i  +i\mu_a\right)
  +    l\left(a- i \sigma_i +i \tilde{\mu}_a \right)
\right]  \ .
\end{split}
}
However this is not yet the magnetic theory discussed in the body of the text.
We further need to use the duality for the $\mathcal{N}=2, \U(N_c=1), N_f=1$ theory.
The magnetic theory is given in \eqref{VVM}, where $M$ is now a $1\times 1$ matrix (a number):
$W=V_{+}V_{-} M$.
At the partition function level this gives the following equality (which holds up to an overall constant term), which is a specialization of
the pentagon identity for quantum dilogarithm:
\es{XYZdual}{
 \int d\sigma \,\ e^{2\pi i b \sigma}
\exp\left[  
    l\left(1-a\pm  i \sigma \right)
    \right] 
    =
\exp\left[  
     l\left(a\pm i b \right) + l\left(1-2a \right)
\right]  \ .
}
After applying \eqref{XYZdual}\footnote{we take $\sigma \to \zeta, \, b\to N_c \hat{b} -\sum_i \sigma_i, 
\, a\to \tilde{N}_c+1-N_f a$ in \eqref{XYZdual}.}, 
the expression \eqref{MagneticPF1-U-all-new} becomes
\es{MagneticPF1-U-all-new2}{
\begin{split}
Z'_{\rm magnetic}& =
{1 \over \tilde{N}_c!}  
\exp\left[ N_f^2 \, l(1-2a) + l\left(1-2(\tilde{N}_c+1-N_f a) \right)  \right]  \\
&\times
\int \prod_{i=1}^{\tilde{N}_c} d\sigma_i 
\prod_{1\le i<j\le \tilde{N}_c} \sinh^2[\pi(\sigma_i-\sigma_j)]  
\\
&
\times 
\exp\left[  
     l\left(\tilde{N}_c+1-N_f a \pm i  N_c \hat{b} \mp i \sum_{i=1}^{\tilde{N}_c}  \sigma_i 
     \right)
\right] 
\\
&
\times
\prod_{i=1}^{\tilde{N}_c} \prod_{a=1}^{N_f}
\exp\left[  
   l\left(a+ i \sigma_i  +i\mu_a\right)
  +    l\left(a- i \sigma_i +i \tilde{\mu}_a \right)
\right]  \ .
\end{split}
}
Again, if we set $\mu_a=\tilde{\mu}_a=0$ this
coincides with the magnetic partition function
\eqref{MagneticPF1} we wrote down in the main text,
under the identification $b=i\hat{b}$.

We discussed above the case of $N_f>N_c$, however
the case of $N_f=N_c$ is similar and simpler, 
so we will not repeat here.

\section{Numerical Tricks}\label{app.numerical}

The evaluation of the our $S^3$ partition function requires
a multi-dimensional integral whose integrands oscillates relatively quickly.
In some cases, we find it numerically more advantageous to convert the multi-dimensional 
integral into a sum of a product of one-dimensional integrals.
Let us illustrate this for the case of $\U(N_c)$ SQCD: the same strategy works in a similar manner
for $\USp(2 N_c)$ and $\SO(N_c)$ SQCD. 

The trick is to use the Weyl character formula
\es{WeylSU}{
\prod_{i<j} 2\sinh \left[ \pi({\sigma_i - \sigma_j}) \right]
=\prod_{i<j} \left( e^{\pi(\sigma_i-\sigma_j) } - e^{\pi(\sigma_j-\sigma_i)}\right)
  = \sum_{\bsigma \in \mathfrak{S}_N} (-1)^{\bsigma}e^{2\pi \sum_i \rho_{\bsigma(i)} \lambda_i} \,,
}
where $\rho:=(\frac{N_c-1}{2}, \frac{N_c-3}{2}, \frac{N_c-5}{2}, \ldots, \frac{-N_c+1}{2})$ is the Weyl vector and $\mathfrak{S}_N$ is the Weyl group ($N$-th symmetric group).

The $\U(N_c)$ electric partition function \eqref{ElectricPF1-U} can then be rewritten as (up to an overall constant factor)
\es{SUreduce_1}{
Z_{\textrm{electric}} 
&={1 \over N_c !} \sum_{\bsigma, \tilde{\bsigma} \in S_{N_c}} (-1)^{\bsigma+\tilde{\bsigma}} \int \prod_{i=1}^{N_c} d \sigma_i
\left[ e^{N_f l(1-a \pm i \sigma_i )} \, e^{2 \pi(\rho_{\bsigma(i)}+\rho_{\tilde{\bsigma}(i)} )\sigma_i} \right] \\
&= \sum_{\bsigma  \in S_{N_c}} (-1)^{\bsigma} \int \prod_{i=1}^{N_c} d \sigma_i 
\left[ e^{N_f l(1-a \pm i \sigma_i)} \, e^{2 \pi(\rho_{\bsigma(i)}+\rho_i) \sigma_i} \right] \\
&= \sum_{\bsigma  \in S_{N_c}} (-1)^{\bsigma} \prod_{i=1}^{N_c} F(\rho_{\bsigma(i)}+\rho_i)  \,,
}
where we defined
\begin{align}
F(x) :=\int \prod_{i=1}^{N_c} d \sigma_i 
\left[ e^{N_f l(1-a \pm i \sigma_i)} \, e^{2 \pi x } \right] \,.
\end{align}
The resulting expression \eqref{SUreduce_1} is now written in terms of $N_c$ one-dimensional integrals. 

\bibliographystyle{nb}
\bibliography{crack}

\end{document}